\documentclass[%
twocolumn,
superscriptaddress,
 amsmath,amssymb,
 aps,pra
]{revtex4}

\usepackage{amsmath}
\usepackage{amssymb}
\usepackage{wasysym}
\usepackage{graphicx}
\usepackage{natbib}
\usepackage{fancyhdr}
\usepackage{mathtools}
\usepackage[usenames,dvipsnames]{xcolor}
\usepackage[utf8]{inputenc}
\usepackage{multirow}
\usepackage{makecell}
\usepackage{ulem} 
\usepackage{xcolor} 
\newcommand\redsout{\bgroup\markoverwith{\textcolor{red}{\rule[0.5ex]{2pt}{0.4pt}}}\ULon} 
\definecolor{amaranth}{rgb}{0.9, 0.17, 0.31}
\begin{document}
\normalem

\title{Electromagnetic theory of Helicoidal Dichroism in reflection from magnetic structures}

\author{Mauro Fanciulli}
\email{mauro.fanciulli@u-cergy.fr}
\affiliation{Universit\'e Paris-Saclay, CEA, CNRS, LIDYL, 91191 Gif-sur-Yvette, France}
\affiliation{Laboratoire de Physique des Mat\'eriaux et Surfaces, CY Cergy Paris Universit\'e, 95031 Cergy-Pontoise, France}

\author{David Bresteau}
\affiliation{Universit\'e Paris-Saclay, CEA, CNRS, LIDYL, 91191 Gif-sur-Yvette, France}

\author{Mekha Vimal}
\affiliation{Universit\'e Paris-Saclay, CEA, CNRS, LIDYL, 91191 Gif-sur-Yvette, France}

\author{Martin Luttmann}
\affiliation{Universit\'e Paris-Saclay, CEA, CNRS, LIDYL, 91191 Gif-sur-Yvette, France}

\author{Maurizio Sacchi}
\affiliation{Sorbonne Universit\'e, CNRS, Institut des NanoSciences de Paris, INSP, F-75005 Paris, France}

\author{Thierry Ruchon}
\email{thierry.ruchon@cea.fr}
\affiliation{Universit\'e Paris-Saclay, CEA, CNRS, LIDYL, 91191 Gif-sur-Yvette, France}

\begin{abstract}

We present the classical electromagnetic theory framework of reflection of a light beam carrying Orbital Angular Momentum (OAM) by an in-plane magnetic structure with generic symmetry. Depending on the magnetization symmetry, we find a change in the OAM content of the reflected beam due to magneto-optic interaction and an asymmetric far-field intensity profile. This leads to three types of Magnetic Helicoidal Dichroism (MHD), observed when switching the OAM of the incoming beam, the magnetization sign, or both. In cases of sufficient symmetries, we establish analytical formulas which link an experimentally accessible MHD signal up to $10\%$ to the Magneto-Optical Kerr Effect (MOKE) constants. Magnetic vortices are particularly enlightening and promising targets, for which we explore the implications of our theory in the joint publication XX XXX XXXXXX.
\end{abstract}
\maketitle

\section{Introduction}
\label{sec:Introduction}
Laguerre-Gaussian light beams, which are a special case of vortex beams (VB) \cite{Padgett2011, Andrews:2008}, carry OAM and show a chiral symmetry: their wavefront appears as a spiral surface, the two forms of which, left-handed or right-handed, are mirror images but not superimposable \cite{Yao2011, Barron2009}. From symmetry principles, one can expect a different response of the two chiralities when interacting with matter presenting no trivial symmetry \cite{CurieJPTA1894}. For instance, such an effect was observed when a VB was sent on chiral molecules adsorbed on a  surface \cite{brullot2016}, or on structured nanopatterns \cite{Mei2016}.  These kinds of differential effects, linked to the properties of light, are generally called birefringence or dichroism, whether they appear on the real or imaginary part of the optical index or reflectivity coefficients. They  may appear due to microscopic, macroscopic, or induced dissymetries in the medium, providing altogether an extremely rich set of investigation tools. 

All these effects have long been identified for light carrying a Spin Angular Momentum (SAM), which is another form of chirality of light beams associated to circular polarization. Discovered by Arago on $\alpha$-Quartz crystals, circular birefringence, also called optical activity, was linked to the symmetry of the macroscopic structure \cite{Arago1811}. It is caused by different real optical indices for circularly polarized light beams with opposite helicities (SAM).  Biot reported four years later that it also has a microscopic origin, reporting its observation in isotropic liquid media \cite{Biot1815}. These seminal discoveries were instrumental towards the discovery by Pasteur of molecular dissymetry \cite{Pasteur}, today called molecular chirality after Kelvin's work \cite{Kelvin1904}. The counterpart in absorption, circular dichroism (CD), was discovered in 1896 by Cotton \cite{cotton1896}.  
Furthermore, the dissymetry can be induced  by external ``forces'', like a magnetic field. Faraday discovered the magnetic circular birefringence in 1846 \cite{Faraday1846} and its counterpart in absorption, Magnetic Circular Dichroism (MCD), which later became accessible \cite{Stephens1974, Ebert:1996}, is now a standard investigation method for magnetic samples, especially in the X-ray spectral range \cite{Sacchi:2001, Funk2005, Laan2014}. All these effects are second-order effects, appearing beyond the first order electric-dipole approximation. However, when the result of the interaction with light is photoionization, first order effects, which appear as an uneven distribution of the photoelectrons on a detector, could be identified, such as for instance PhotoElectron Circular Dichroism (PECD) \cite{Ritchie:1976,Powis2008}, or Circular Dichroism in Angular Distribution (CDAD) on surfaces \cite{schoenhense1991}.

As for OAM-dependent light-matter interactions, which we may call Helicoidal Dichroisms (HD) \footnote{Ref.~\cite{Veenendaal2007} calls it ``OAM-induced dichroism'', while Ref.~\cite{brullot2016} uses ``Helical Dichroism''.}, a few of them have already been reported in different contexts, as reviewed in Ref.~\cite{shen2019}. To list only a few examples, one could consider the measurements of the OAM of light beams as a kind of HD, may it be used in the framework of classical \cite{Lavery2012} or quantum light \cite{Kulkarni2017,Berkhout2010}. These schemes use either diffraction on non symetrical apertures (e.g. triangular slit), interferences with a beam of different symmetry (e.g. a VB with a plain beam), or modes converters using birefringent prisms. But none of these schemes is currently of spectroscopic wide interest: matter is here only used to alter the mode content of the beam, without any consideration upon the physics of light-matter interaction. 

The question of spectroscopic applications first arises when matter is left in an excited state. A recent specific review dedicated to the interaction of twisted light with atoms is available in Ref.~\cite{Babiker2018}. Briefly, through electric-dipole transitions, twisted light beams do not couple differentially to the internal degrees of freedom of the atomic or molecular system, \emph{i.e.} the electronic ones, but can act on its external ones \cite{Babiker2002, Giammanco2017}. This last behavior led to important developments, enriching the scope of techniques available for manipulation and cooling of atoms. The first order of perturbation sensitive to the OAM linked to internal degrees of freedom is the electric quadrupolar one, as recently demonstrated experimentally in ultra-cold trapped ions \cite{Schmiegelow2016,SolyanikGorgone2019}, or theoretically in oriented chiral molecular ensembles \cite{Forbes:2018}. Such a result also holds for  bulk magnetic materials \cite{Veenendaal2007}. Interestingly, these general conclusions are modified when an atomic system gets ionized through the interaction with a very high intensity beam ($\gtrsim 10^{20} \text{W/cm}^2$): in this context new selection rules were proposed within the electric dipole approximation \cite{picon2010, Waetzel2016,FrankeArnold2017, Giri2020}, providing an analog to PECD and CDAD. However, all these spectroscopic HDs remain experimentally extremely challenging nowadays, if possible at all. The difficulty ultimately relies on the necessity for the system to ``see'' altogether, a significant twist of the wave front over its dimensions, and a high enough intensity. 
When systems are larger than atoms or molecules, these conditions may be less demanding. An important exemple is the control of Bose-Einstein Condensates with VB \cite{Molina-Terriza2007}. But classical objects could also be considered. For instance, an effect was predicted with nanodots \cite{Quinteiro2010}, and an OAM-dependent plasmonic coupling between SAM and OAM was reported when light is sent through nanoholes \cite{zambranaNC2014}.

From this brief and selective overview, although many applications emerged using VB and OAM beams \cite{shen2019,Yao2011},  it appears that the picture of anistropic effects involving VB remains incomplete as compared to circularly polarized beams. It is even more apparent when considering reflective geometries for which a wealth of magneto-optical Kerr effect (MOKE) \cite{Oppeneer2001, Ebert:1996, Arregi:2016, Yamamoto2017} have been identified for different polarization/magnetization combinations which do not find their counterpart for VB. These MOKE effects are particularly enlightning in the Extreme Ultra Violet (XUV) spectral range (50-150 eV), where 3$p$ edges of many magnetic material are found \cite{Sacchi:1998b, Grychtol:2010, ferrari2016}, and in the soft X-ray region (600-900 eV), typical for 2$p$ edges \cite{Kao:1990, Sacchi:1998, Sacchi:1999}.

In this article, we contribute to filling the picture by explicitly predicting the existence of a phenomenon analogous to MCD, observable with beams carrying OAM instead of SAM.  Its value is in the 10\% range, comparable to other MOKE effects. Combined with the recent availability of XUV VB both on Free-Electron Laser sources \cite{Rebernik-Ribic2014, Ribic2017, Lee:2019} and High Harmonic Sources \cite{Gariepy2014,Geneaux2016,Gauthier2017,Kong2017,Rego2016,Dorney2018,Rego2019}, it should make it measurable rapidly. 
We consider structures with sizes comparable to a standard beam focus (100\,nm-few $\mu$m width), and materials which exist at ambient temperature. This lifts the above mentioned strong requirements, making MHD a promising spectroscopic tool. 
For simplicity, only magnetization with constant magnitude, not radially dependent and with in-plane components is considered, but an extension to more general cases can be readily achieved. 
We will derive the analytical expressions of MHD for reflection of beams carrying OAM in the three different cases of switching the OAM sign, the magnetization sign or both. 
We find that for targets with non homogeneous magnetization MHD is always present when the reflected beam profile is spatially resolved, which we indicate as ``differential'' MHD. This is similar to what has been shown for the case of resonant X-ray scattering of light carrying SAM \cite{chauleau2018} or OAM \cite{Veenendaal2015}. 
Furthermore, while MHD also depends on the polarization state of the incident light, its observation does not require any polarimetric analysis, which is convenient especially for the XUV spectral range. This possibility is explored in detail for the special case of a magnetic vortex in the joint publication Ref.~\cite{MHDprl}. We implement our model in numerical calculations, the details of which are reported in Appendix~\ref{app:numerical}.

This paper is organized as follows.
In Section \ref{sec:model} we present the analytical model for the input OAM beam and a generic magnetic structure. The two special cases of a magnetic vortex and of two antiparallel magnetic domains are considered explicitly.
In Section \ref{sec:selectionrules} we calculate the characteristics of the reflected light beam by the magnetic structure in the near field, finding the rules for the modification of the OAM.
In Section \ref{sec:hd} we propagate the beam to the far field, evaluate the expression for the intensity and find the equations describing the MHD.
Finally, discussion and conclusions are presented in Section~\ref{sec:conclusion}.\\

\section{Model}\label{sec:model}
In this Section we present the analytical framework used in our model to describe the Laguerre-Gaussian (LG) beam and the magnetic structure.
\subsection{Beam propagation and decomposition on the Laguerre-Gaussian basis}
\begin{figure}[htbp]
\includegraphics[width=0.48\textwidth]{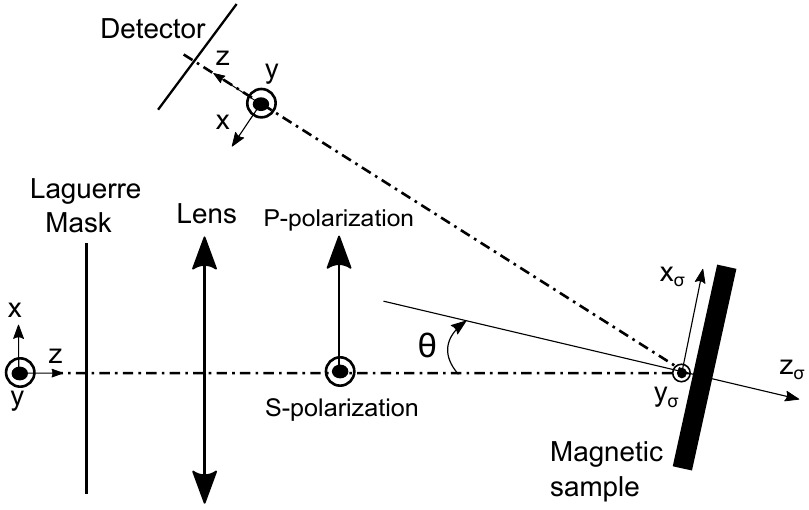}
\caption{\label{fig:manip}Sketch of the beam path for the reflection on the magnetic target.}
\end{figure}

We consider sufficiently loose focusing conditions so that the paraxial equation for beam propagation is valid. We start with a collimated Gaussian beam and propagate it in several steps using the Fresnel integral. The beam path is shown in Fig.~\ref{fig:manip}. It first goes through a phase mask, such as a spiral staircase, which imparts OAM to the beam. Then it propagates to a lens, and it is focused on the sample. After reflection, the beam is again propagated to the far field, where the detector is placed.
In order to be able to interpret the results as angular momenta transfers, we use the LG basis to analyse the interaction of an optical vortex with a magnetic structure. The LG basis is a family of solutions of the paraxial equation forming a complete basis, indexed by the two integers $(\ell, \rho)$, with $\ell$ being the azimuthal number, positive or negative, and $\rho$ the radial number, positive.
The complex electric field of a given LG mode (focus at $z=0$) reads \cite{Allen1992}: 
\begin{multline}\label{eq:LGBeam}
\vec{E}_{in}=C_\rho^{|\ell|}\frac{1}{w(z)}\left(\frac{r\sqrt{2}}{w(z)}\right)^{|\ell|}\cdot L_\rho^{|\ell|}\!\left(\frac{2r^2}{w^2(z)}\right)\cdot\\
 \cdot e^{-\frac{r^2}{w^2(z)}}e^{-ik\frac{r^2}{2R(z)}}e^{i\ell\phi}e^{i(\omega t-kz)}e^{i(2\rho+|\ell|+1)\gamma(z)}
\begin{pmatrix}
\epsilon_p\\
\epsilon_s
\end{pmatrix}\,,
\end{multline}
where $C_\rho^{|\ell|}$ is a normalization constant specific for the basis $(\ell,\rho)$, $(r,\phi)$ are the polar coordinates, $L_\rho^{|\ell|}$ represents the Laguerre polynomial, $\omega$ the angular frequency, $k$ the wave vector along the propagation direction $z$. 
The vector $(\epsilon_{p},\epsilon_{s})$ represents the polarization state of the beam; notably, $(1,0)$, $(0,1)$ and $(1,\pm i)$ correspond to linearly $P$-polarized, linearly $S$-polarized and positive and negative circularly polarized beam, respectively. The Gouy phase $\gamma(z)$, the beam width $w(z)$ and the radius of curvature $R(z)$ are respectively defined as:
\begin{subequations}
\begin{align}
				\gamma(z)&=\arctan\left(\frac{z}{z_R}\right)\\
				w(z)&=w_0\sqrt{1+\left(\frac{z}{z_R}\right)^2}\\
				R(z)&=z\left[1+\left(\frac{z_R}{z}\right)^2\right]
\end{align}
\end{subequations}
with $z_R=\pi w_0^2/\lambda$ being the Rayleigh range and $w_0$ the waist of the beam.
For shortness, we rewrite Eq.~\eqref{eq:LGBeam} as:
\begin{equation}
\vec{E}_{in}=A_\rho^{|\ell|}e^{i\varphi_0}e^{i\ell\phi}\begin{pmatrix}
\epsilon_p\\
\epsilon_s
\end{pmatrix}
\label{eq:LGBeam2}
\end{equation}
where we introduced $A_\rho^{|\ell|}(r,z)=C_\rho^{|\ell|}\frac{1}{w(z)}\left(\frac{r\sqrt{2}}{w(z)}\right)^{|\ell|}L_\rho^{|\ell|}\!\left(\frac{2r^2}{w^2(z)}\right)e^{-\frac{r^2}{w^2(z)}}e^{-ik\frac{r^2}{2R(z)}}$ and $\varphi_0(t,z)=\omega t-kz+(2\rho+|\ell|+1)\gamma(z)$.

\subsection{Model for the optical properties of a magnetic structure}\label{VortexModel}
We model our sample as a magnetic material of typical extension $R_0$ deposited on a non-magnetic surface. We will use the $\sigma$ subscript for the sample's surface frame, described in cartesian coordinates $(x_\sigma,y_\sigma)$ and polar coordinates $(r_\sigma,\phi_\sigma)$.
For the sake of simplicity of the analytical derivations, we will consider a structure that is perfectly flat and is larger than the incident beam ($R_0>w_0$), so that geometrical and diffraction effects need not be taken into account. Also, we will consider structures with constant magnetization magnitude and no radial dependence, so that the only variation is due to direction change with azimuthal dependence $\phi_\sigma$. For a better description of the structure in numerical calculations see Appendix~\ref{app:numerical}.

The reflection of the light beam is modeled by the reflectivity matrix $\mathbf{R}$, for which $\vec{E}_{out}=\mathbf{R}\vec{E}_{in}$. When a light beam is reflected off a magnetic surface, the standard Fresnel reflectivity coefficients for the S and P polarizations, denoted $r_{ss}$ and $r_{pp}$, are complemented by magnetization-dependent terms $r_{ps}^l$ and  $r_0^t$ describing the Magneto-Optical Kerr Effect (MOKE) \cite{Oppeneer2001, Piovera2013}.
The first coefficient couples the $S$ and $P$ polarizations in presence of a longitudinal magnetization, while the second acts only on the $P$ polarization when there is a transverse component of the magnetization.

For simplicity, we limit ourselves to the case of in-plane magnetization only, but the model can be readily extended to the out-of-plane magnetization component (i.e. polar MOKE component) as well. We also limit ourselves to the MOKE terms linear with the magnetization \footnote{However, arguments similar to those of Section~\ref{sec:tilt} would show that the quadratic terms do not lead to a dependence of the reflected pattern on the sign of the OAM.}. With all these restrictions, we model the reflection matrix as \cite{Piovera2013}:
\begin{equation}
\mathbf{R}(\phi_\sigma)=\begin{pmatrix}
r_{pp}\cdot\left[1+ r_0^t \cdot m_t(\phi_\sigma)\right]& r_{ps}^l \cdot m_l(\phi_\sigma) \\
- r_{ps}^l\cdot m_l(\phi_\sigma) & r_{ss}
\end{pmatrix}
\label{eq:reflectionMatrixA}
\end{equation}
where we have explicitly indicated the dependence of the matrix on the azimuthal location on the sample.
$m_t$ and $m_l$ are defined as $m_t=M_t/M_S$ and $m_l=M_l/M_S$, where $M_S$ is the saturation magnetization of the sample and $M_t$ and $M_l$ are the magnetization components along the transverse (perpendicular to the scattering plane, i.e. along $y_\sigma$) and longitudinal (parallel to the scattering plane, i.e. along $x_\sigma$) directions with respect to the scattering plane in the sample frame (see Fig.~\ref{fig:manip}).
All the four reflectivity coefficients are complex quantities, 
which we consider constant over the structure.

Now we need to model the azimuthal dependence of the magnetization.
The formalism of Eq.~\eqref{eq:reflectionMatrixA} requires to express the magnetization in its longitudinal and transverse components, $\vec{m_l}=m_l\hat{x_\sigma}$ and $\vec{m_t}=m_t\hat{y_\sigma}$. To take advantage of the symmetries of the problem, we expand the angular part on the standard basis functions, $\frac{1}{\sqrt{2\pi}}e^{i n \phi_\sigma}$.
For any function sufficiently regular, the magnetization $m_*$ (with $*=l,t$) can thus be written as: 
\begin{equation}
m_*(\phi_\sigma)=
\sum_{n=-\infty}^{+\infty}m_{*,n}e^{in\phi_\sigma}
\label{eq:decomp}
\end{equation}
with complex decomposition coefficients:
\begin{equation}
m_{*,n}=\frac{1}{2\pi}\int_0^{2\pi}m_*(\phi_\sigma)e^{-in\phi_\sigma}d\phi_\sigma
\label{eq:decomp2}
\end{equation}
We notice here that since $m_*$ is a real quantity, we have the property $m_{*,-n}=\overline{m_{*,n}}$. Now we can rewrite the reflection matrix as:
\begin{equation}
\mathbf{R}(\phi_\sigma)=\begin{pmatrix}
r_{pp}& 0\\
0 & r_{ss}
\end{pmatrix}+\sum_{n}
\begin{pmatrix}
r_{pp} r_0^t m_{t,n} e^{in\phi_\sigma} & r_{ps}^lm_{l,n}e^{in\phi_\sigma} \\
- r_{ps}^lm_{l,n}e^{in\phi_\sigma}& 0
\end{pmatrix}
\label{eq:reflectionMatrix2}
\end{equation}
where we have separated the magnetization dependent and independent parts.

\subsubsection*{Magnetic structures with high symmetries}
There is \emph{a priori} no restriction on the span of the $n$ values in the decomposition, 
and different specific geometries will differ by their decomposition coefficients.
We analyze in more details some cases of high symmetry that are relevant for magnetization structure, as defined in the following. 
For each component $m_*$ of the magnetization, we consider even or odd symmetries with respect to (w.r.t.) the $x_\sigma$ and $y_\sigma$ axes. Each of these four cases has specific consequences on the properties of the decomposition coefficients, and in particular on their parity. These properties are demonstrated in Appendix~\ref{AppendixSym}, and summarized in Table~\ref{tablesym}.
Given the two components of the magnetization, this restriction leads to 16 different cases of symmetries. In order to further simplify, we consider only the cases of magnetic structures having longitudinal and transverse components with same parities for their decomposition coefficients. This additional restriction leads to 8 cases of symmetry, which fall within the same formalism described in Section~\ref{sec:example}.
\begin{table}[ht]
\centering
\begin{tabular}{cc|p{2.5cm}|p{2.5cm}|}
\cline{3-4}
& & \multicolumn{2}{ c| }{Symmetry w.r.t. $y_\sigma$} \\ \cline{3-4}
& & \quad\quad\quad even & \quad\quad\quad odd \\ \cline{1-4}
\multicolumn{1}{ |c  }{\multirow{2}{*}{\begin{tabular}{@{}c@{}}\\Symmetry\\ w.r.t. $x_\sigma$\end{tabular}} } &
\multicolumn{1}{ |c| }{even} & \makecell{$n$ \textbf{even} only\\$m_{*,n}$ \textbf{real}\\$m_{*,n}=m_{*,-n}$}& \makecell{$n$ \textbf{odd} only\\$m_{*,n}$ \textbf{real}\\$m_{*,n}=m_{*,-n}$}     \\ \cline{2-4}
\multicolumn{1}{ |c  }{}                      &
\multicolumn{1}{ |c| }{odd} & \makecell{$n$ \textbf{odd} only\\$m_{*,n}$ \textbf{imaginary}\\$m_{*,n}=-m_{*,-n}$} & \makecell{$n$ \textbf{even} only\\$m_{*,n}$ \textbf{imaginary}\\$m_{*,n}=-m_{*,-n}$}      \\ \cline{1-4}
\end{tabular}
\caption{Properties of the decomposition coefficients for a particular component of the magnetization, $m_*=m_l$ or $m_*=m_t$, depending on its parity with respect to the $x_\sigma$ and $y_\sigma$ axes.}
\label{tablesym}\end{table}

\subsubsection*{Two examples of magnetic structures}
\begin{figure}[ht!]
\includegraphics[width=0.45\textwidth]{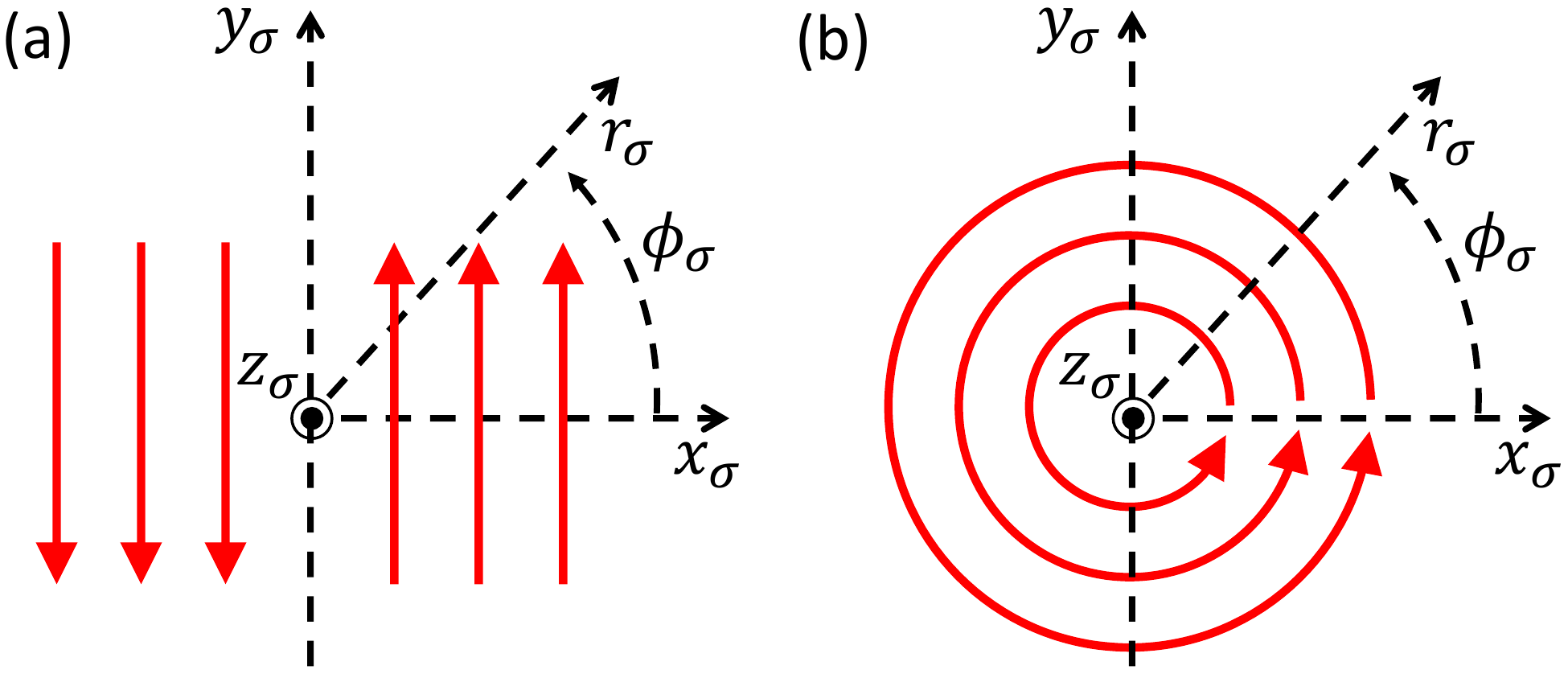}
\caption{\label{fig:examples}Examples of magnetic samples with cartesian and polar coordinate systems: (a) two homogeneous and antiparallel magnetic domains; (b) counterclockwise magnetic vortex.}
\end{figure}

We provide now two prototypical examples of magnetic configurations, depicted in Fig.~\ref{fig:examples}. In the first one [Fig.~\ref{fig:examples}(a)] the sample separates into two magnetically homogeneous domains of equal size, aligned antiparallel to each other. The second case [Fig.~\ref{fig:examples}(b)] consists of a magnetic vortex with counterclockwise circulation of the magnetization.

In the case of two antiparallel domains [Fig.~\ref{fig:examples}(a)], the magnetization can be represented as:
\begin{subequations}
\begin{align}
m_l (\phi_\sigma)&=0\\
m_t (\phi_\sigma)&=m_0\text{ sign}\left(\cos\phi_\sigma\right),
\end{align}
\end{subequations}
with $m_0$ being the constant magnitude of the magnetization normalized by $M_S$.
In this case, the symmetry of $m_t$ w.r.t. $\left(x_\sigma, y_\sigma\right)$ is $\left(\text{even},\text{odd}\right)$. To find the corresponding coefficients, integration of Eq.~\ref{eq:decomp2} gives (see Appendix~\ref{app:magneticexamples} for the calculation):
\begin{equation}
m_{t,n} = \left\{\begin{array}{ll}
        0, & \text{for }n\text{ even}\\
       \frac{2}{n\pi} \;i^{n-1}\; m_0, & \text{for }n\text{ odd}
        \end{array}\right.
        \label{decompositiondomainwall}
\end{equation}
In the case of the magnetic vortex [Fig.~\ref{fig:examples}(b)] only two coefficients of the decomposition for both components are non zero, corresponding to $n=\pm 1$:
\begin{subequations}
\begin{align}
m_l (\phi_\sigma)&=\frac{m_0}{2i}e^{-i\phi_\sigma}-\frac{m_0}{2i}e^{i\phi_\sigma}=-m_0\sin \phi_\sigma\\
m_t (\phi_\sigma)&=\frac{m_0}{2}e^{-i\phi_\sigma}+\frac{m_0}{2}e^{i\phi_\sigma}=m_0\cos \phi_\sigma
\end{align}
\end{subequations}
In this case, the symmetry of $m_l$ and $m_t$ w.r.t. $\left(x_\sigma, y_\sigma\right)$ is respectively $\left(\text{odd},\text{even}\right)$ and $\left(\text{even},\text{odd}\right)$. The integration in Eq. \ref{eq:decomp2} now gives (Appendix~\ref{app:magneticexamples}):
\begin{equation}
m_{*,n} = \left\{\begin{array}{ll}
        0, & \text{for  }n\neq \pm 1\\
       \frac{i m_0}{2} n, & \text{for } n =\pm 1, *=l\\
        \frac{ m_0}{2}, & \text{for } n =\pm 1, *=t
        \end{array}\right.
        \label{decompositionVortex}
\end{equation}\\

\section{Modification of the OAM by reflection on a magnetic structure}\label{sec:selectionrules}
We consider the axis of the OAM beam to pass through the center of the magnetic target. 
It should be noted that realistic sizes of both magnetic structures and OAM beams can range from hundreds of nanometers to several micrometers. Therefore an experimental implementation can be achieved, with accurate control of the sample holder and beam steering mirrors. 
By using Eqs.~\eqref{eq:LGBeam2} and~\eqref{eq:reflectionMatrix2}, we calculate the outgoing electric field $\vec{E}_{out}=\mathbf{R}\vec{E}_{in}$, separating the result into two terms corresponding to the non-magnetic ($\vec{E}_{out}^{m=0}$) and magnetic ($\vec{E}_{out}^{m}$) interaction, $\vec{E}_{out}=\vec{E}_{out}^{m=0}+\vec{E}_{out}^{m}$:
\begin{widetext}
\begin{subequations}\label{eq:Eout}
\begin{align}
\vec{E}_{out}^{m=0}&=A_\rho^{|\ell|}(r,0)e^{i\varphi_0}e^{i\ell\phi}\begin{pmatrix}
\epsilon_p r_{pp}\\
\epsilon_s r_{ss}
\end{pmatrix}\label{Em0}\\
\vec{E}_{out}^{m}&=A_\rho^{|\ell|}(r,0)e^{i\varphi_0}\sum_{n}e^{i\left(\ell\phi+n\phi_\sigma\right)}\begin{pmatrix}
\epsilon_p r_{pp} r_0^t m_{t,n}+\epsilon_s r_{ps}^l m_{l,n}\\
-\epsilon_p r_{ps}^l m_{l,n}
\end{pmatrix}\label{Em}
\end{align}
\end{subequations}
\end{widetext}

We consider that there are no homogeneous magnetization terms: $m_{t,0}=m_{l,0}=0$, since it will considerably simplify the derivation without much loss of generality. In this way we can express the total field as:
\begin{equation}\label{eq:Eout2}
\vec{E}_{out}=\sum_{n}\vec{E}_{n,\ell}=A_\rho^{|\ell|}(r,0)e^{i\varphi_0}\sum_{n}e^{i\left(\ell\phi+n\phi_\sigma\right)}
\begin{pmatrix}
\alpha_{n,m}^{x}\\
\alpha_{n,m}^{y}
\end{pmatrix}
\end{equation}
where we defined the complex quantities $\alpha_{n,m}^{x}$ and $\alpha_{n,m}^{y}$ for $n\neq 0$ and $\alpha_{0,m}^{x}$ and $\alpha_{0,m}^{y}$ as:
\begin{equation}
    \begin{aligned}
    \alpha_{0,m}^{x}&=\epsilon_p r_{pp}\\
    \alpha_{n,m}^{x}&=m\left(\epsilon_p r_{pp} r_{0}^t m_{t,n}+\epsilon_s r_{ps}^l m_{l,n}\right)\\
    \alpha_{0,m}^{y}&=\epsilon_s r_{ss}\\
    \alpha_{n,m}^{y}&=m\left(-
\epsilon_p r_{ps}^l m_{l,n}\right)\\
    \end{aligned}\label{alphas}
\end{equation}
The index $m=\pm1$ is introduced in order to explicitly describe the reversal of the magnetization direction in Section~\ref{sec:hd}, and acts only for $n\neq0$.

\subsection{Normal incidence}\label{General}
\begin{figure}[!htp]
\centering
\includegraphics[width=.98\columnwidth]{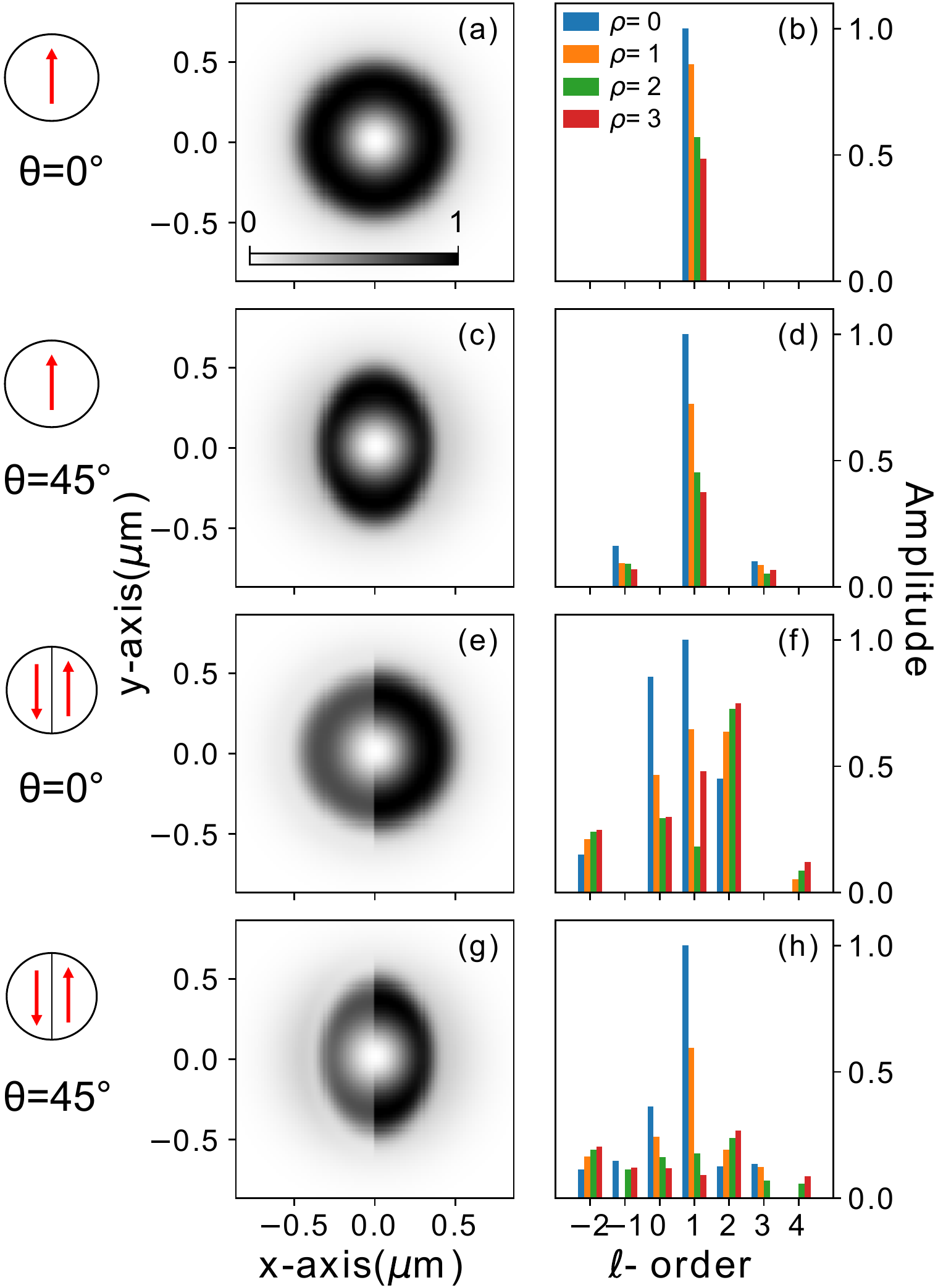}
\caption{
(left) Near field intensity profiles and (right) decomposition on the LG-mode basis of a incoming beam with $\ell=1$ reflected by a magnetic dot of radius $500$~nm in four different cases. (a),(b) constant magnetization with incidence angle $\theta=0^\circ$ and (c),(d) with $\theta=45^\circ$; (e),(f) two antiparallel magnetic domains as in Fig.~\ref{fig:examples}(a) with incidence angle $\theta=0^\circ$ and (g),(h) with $\theta=45^\circ$. The focal spot has a size comparable to that of the magnetic dot. The bar plots values below 4\% of the maximum have been forced to 0.  Details of the numerical calculations are reported in Appendix~\ref{app:numerical}.
\label{FigTilt_NoTilt}}
 \end{figure}

We consider here the case of normal incidence of the beam. It is important to point out that Eq.~\eqref{eq:reflectionMatrixA} is written with respect to a well defined scattering plane, which means off-normal incidence. Therefore this is a simplification that will allow us to better understand the effect of the magnetic structure on the reflection and to separate it from geometrical effects. Pragmatically, this could be seen as the situation where the incidence angle is very close to normal incidence. We will consider the effect of a tilted target in Sec.\ref{sec:tilt}, and we will see in which conditions it is possible to distinguish between the two effects, the trivial geometrical effect and MHD.
For normal incidence, the beam and the target share the same polar coordinates, therefore we set $\phi_\sigma=\phi$ and $r_\sigma=r$ in Eq.~\eqref{eq:Eout}.
The field after reflection $\vec{E}_{out}$ is a superposition of different modes. The nonmagnetic term is an OAM beam of the same order as the incoming one. Its polarization will be different from the incoming one due to the different values or $r_{pp}$ and $r_{ss}$ [Eq.~\eqref{Em0}], corresponding to the Kerr effect. 
The magnetic term is more interesting [Eq.~\eqref{Em}], and two observations can be made at this point:
\begin{enumerate}
\item The beam is no longer a pure LG mode, since the $|\ell|$ power of $r$ in $A$ and the azimuthal phase no longer match. This leads to the appearance of radial modes $\rho$ different from the incoming one. This effect is already documented in linear processes and was lately rationalized for low order non linear effects \cite{Buono:2020}.
	\item 
	The refleceted beam has a different azimuthal mode population with respect to the incoming one.
	In particular, the incoming OAM of order $\ell$ will give rise to all the possible orders $\ell+n$ for every $n$ belonging to the decomposition of Eq.\eqref{eq:decomp}. 
\end{enumerate}
For example, for the case of a magnetic vortex only the coefficients of the decomposition corresponding to $n=\pm1$ are non zero, according to Eq.~\eqref{decompositionVortex}. Therefore the interaction of an OAM of order $\ell$ with a magnetic vortex results in the population of the $\ell\pm 1$ modes. This situation is particularly suitable for the study of MHD, and is described in detail in the joint publication Ref.~\cite{MHDprl}. As another example, in Fig.~\ref{FigTilt_NoTilt} we compare the case of a magnetic dot with constant magnetization and a dot with two antiparallel domains as in Fig.~\ref{fig:examples}(a), for an incoming beam with $\ell=1$. The size of the dot is chosen similar to the beam waist at focus (radius of 500 nm). The details of the numerical calculations are described in Appendix~\ref{app:numerical}. The near field intensity profile and the decomposition on the LG basis are shown in Fig.~\ref{FigTilt_NoTilt}, panels (a) and (b) for the case of constant magnetization, while the corresponding ones for the two domains are shown in panels (e) and (f). While there is no modification to the population of $\ell$ modes in the reflected beam in the case of constant magnetization, in panel (f) we can clearly see the population of $\ell+n$ modes with only $n$ odd terms. In both cases we find a rich set of radial modes $\rho$ due to the finite size of the target.

It is useful to consider separately the $S$ and $P$ components of the incoming beam. For the $S$ part we set $\epsilon_p=0$. In this case $\vec{E}_{out}^{m=0}$ is along the $y$ direction and $\vec{E}_{out}^{m}$ along the $x$ direction. Taking the square of their sum to get the intensity leads to no cross terms. Instead, for the $P$ component of the incoming field we set $\epsilon_s=0$, and we find that $\vec{E}_{out}^{m=0}$ is along the $x$ direction while $\vec{E}_{out}^{m}$ has both $x$ and $y$ components. Taking the square of their sum leads to a cross term, which is at the origin of the dichroism effect that will be discussed in more detail in Sec.\ref{sec:hd}.

\subsection{Arbitrary incidence angle
}\label{sec:tilt}
We now generalize the model to describe the reflection of an OAM beam at an arbitrary angle of incidence, since in practical situations the sample will be rotated with respect to the incoming beam, in order to be able to collect the reflected beam. This has two effects on the formalism presented before. One is that the reflectivity coefficients will depend on the angle. This can be trivially taken into account and exploited to choose favourable conditions, in which the magneto optical constants will be large compared to the regular Fresnel reflection coefficients. For instance, one may choose the Brewster angle to maximise the magnetic contribution to the reflected intensity, as in transverse MOKE experiments using P-polarized light, with and without polarization analysis \cite{Oblak:2017}.
The other effect is a change in the geometry, since the polar coordinates of the beam and the target are no longer the same.
For instance, if we consider a magnetic vortex where the magnetization shows a circular pattern, from the point of view of the beam it will appear as an ellipse. This will give rise to a trivial dichroism, which might have different symmetry than the magnetic HD of interest. In order to study this effect, we consider a rotation of the target by an angle $\theta$ as depicted in Fig.~\ref{fig:manip}. In this case, the relationship between the cartesian coordinates in the sample frame $(x_\sigma,y_\sigma)$ and in the beam frame $(x,y)$ is:
\begin{subequations}
\begin{align}
	x_\sigma&=\frac{x}{\cos\theta}\label{change1}\\
	y_\sigma&=y\label{change2}.
\end{align}
\end{subequations}
Upon trigonometric inspection and defining the function $g(\phi,\theta)=1/\sqrt{1-\sin^2\phi\sin^2\theta}$ one finds the following relations [see Appendix~\ref{sec:trigonometry}]:
\begin{subequations}\label{eq:trigonometry}
\begin{align}
\sin \phi_\sigma&=\sin\phi\cdot g(\phi,\theta)\cdot \cos\theta\\
\cos \phi_\sigma&=\cos\phi\cdot g(\phi,\theta)\\
e^{i\phi_\sigma}&=g(\phi,\theta)\left[\frac{1+\cos\theta}{2}e^{i\phi}+\frac{1-\cos\theta}{2}e^{-i\phi}\right]\label{phi_p}\\
r_\sigma&=\frac{r}{g(\phi,\theta)\cdot\cos\theta}
\end{align}
\end{subequations}
As expected, for $\theta=0$ we retrieve the coincidence of the polar coordinates of beam and sample. The azimuthal dependence of the functions of our chosen basis is $e^{in\phi_\sigma}$. With Eq.~\eqref{phi_p} we get 
\begin{equation}
    \begin{aligned}
     e^{in\phi_\sigma}&=g(\phi,\theta)^n\cdot e^{in\phi}\cdot\\
     &\cdot\sum_{n_1=0}^n C_{n_1}^n\left(\frac{1+\cos\theta}{2}\right)^{n_1}\cdot\left(\frac{1-\cos\theta}{2}\right)^{n-n_1}e^{i2n_1\phi}\label{phi_2}   
    \end{aligned}
\end{equation}
where $C_{n_1}^n$ is the number of combinations of $n_1$ elements from $n$. 
This expression should be inserted in Eq. \eqref{Em}. Without expressing it fully, we see right away that the phases previously reading $\ell\phi+n\phi$ now become $\ell\phi+(n+2n_1)\phi$. For instance, in the case of the magnetic configurations depicted in Fig.~\ref{fig:examples} where only odd coefficients are present, the even coefficients remain zero when considering a tilted target.  This is clearly shown in Fig.~\ref{FigTilt_NoTilt}(c),(d), where we chose $\theta=45^\circ$.

Additionally, the radius $r_\sigma$ and the $g$-function also become function of $\phi$, with a $\sin^2 \phi$ dependence. Thus they are symmetric with respect to $\phi=\pi$ and $\phi=\pi/2$ and their decomposition yields only even coefficients. We notice that if the incoming beam carries an odd value of the OAM, the phase-dependent term will appear in the odd LG modes, while the radial-dependent terms will populate the even LG modes. Conversely, if the incoming beam carries an even value of OAM, phase-dependent terms will populate the even modes and radial-dependent terms the odd modes.
Therefore their influence can always be separated. In particular, we come to the conclusion that the magnetic terms will show up in the LG components of opposite parity compared to the incoming beam, while the non magnetic terms populate modes of the same parity, which modifies the observation 2 of the previous section where all the $\ell+n$ modes are populated by the magnetic term and only $\ell$ by the nonmagnetic one. This case is illustrated in Fig.~\ref{FigTilt_NoTilt}(g),(h) (see Appendix~\ref{app:numerical} for further details).\\

\section{Helicoidal dichroism in the far field}\label{sec:hd}
So far we focused on the structure and mode content of the field right after reflection by the magnetic object. Here we show how the far field profile of the beam is affected, leading to what we call differential helicoidal dichroism, i.e. a difference in intensity profiles upon switching the sign of the OAM, of the magnetization, or of both. Here ``differential'' is meant as a MHD that occurs at every single point in the image in the far field. However, the integration over space in the image leads to constant total intensity, independent on the sign of the OAM or of the magnetization. 
This is qualitatively different from MCD where the difference persists also after spatial integration of the scattered intensity. In this respect, MHD is similar to CDAD of photoemission, which is also a spatially differential effect with no dichroic signal of the spatially integrated intensity. 

In the following we will describe analytically only the case of normal reflection, and in Section~\ref{sec:conclusion} we will discuss how a tilted sample can lead in practice to favorable conditions, along the lines of what was discussed in Section~\ref{sec:tilt}.

\subsection{Propagation of the reflected field from the focus to the far field}
In order to propagate $\vec{E}_{out}$ to the far field, we make use of the Fresnel operator.
From now on, we refer to the far field on the screen in Fig.~\ref{fig:manip} with the coordinate system $(r,\phi,z)$, and we indicate with the prime the field right after reflection, i.e. the one from Eq.~\eqref{eq:Eout2}.
For a function $E(r,\phi,0)$ with separable variables [as it is in our case of Eq.~\eqref{eq:Eout2}], the Fresnel propagation equation in cylindrical coordinates reads \cite{Born:1999}: 
\begin{equation}
\begin{aligned}
E(t,r,\phi,z)&=
\frac {e^{i\varphi_0(t,z)}}{i\lambda z}e^{\frac {ik}{2z}r^2}\int_{0 }^{2\pi}d\phi'\\
&\int_0^{+\infty }r'dr'\;E(r',\phi',0)e^{-\frac {ikrr'}{z}\cos\left(\phi-\phi'\right)}\label{eq:Fresnel2}
\end{aligned}
\end{equation}
The expression in Eq.~\eqref{eq:Fresnel2} has the form of a specialized Fourier transform, or a generalized Hankel transform.
Therefore we propagate the field of Eq.~\eqref{eq:Eout2} with Eq.~\eqref{eq:Fresnel2}.
We study what happens to the propagation of any term $E_{n,\ell,m}^{x,y}(r',\phi',0)=A_\rho^{|\ell|}(r',0)e^{i\left(\ell+n\right)\phi'}\alpha_{n,m}^{x,y}$, and the final result will be given by the  sum over $n$. We have:
\begin{widetext}
\begin{equation}
\begin{aligned}
E_{n,\ell,m}^{x,y}(t,r,\phi,z)&=\frac {e^{i\varphi_0\left(t,z\right)}}{i\lambda z}e^{\frac {ik}{2z}r^2}\alpha_{n,m}^{x,y}\int_0^{+\infty }r'dr'A_\rho^{|\ell|}(r',0)\int_{0 }^{2\pi}d\phi'\;e^{i(\ell+n)\phi'}e^{-\frac {ikrr'}{z}\cos\left(\phi-\phi'\right)}\\
&=\frac {e^{i\varphi_0\left(t,z\right)}}{i\lambda z}e^{\frac {ik}{2z}r^2}\alpha_{n,m}^{x,y}e^{i(\ell+n)\phi}\int_0^{+\infty }r'dr'A_\rho^{|\ell|}(r',0)\int_{0 }^{2\pi}d\phi''\; e^{i(l+n)\phi''}e^{-\frac {ikrr'}{z}\cos\phi''}
\label{eq:Fresnel3}
\end{aligned}
\end{equation}
with the substitution $\phi''=\phi'-\phi$. The last integration is just the definition of the Bessel function $J_{\ell+n}\left(\frac {krr'}{z}\right)$ multiplied by a factor $2\pi/i^{l+n}$ \cite{Born:1999}. We keep implicit the trivial dependence on time in $\varphi_0(z)=\varphi_0(t,z)=\omega t-kz+(2\rho+|\ell|+1)\gamma(z)$, while we write explicitly the expression of $A_\rho^{|\ell|}(r',0)=C_\rho^{|\ell|}\frac{1}{w_0}\left(\frac{r'\sqrt{2}}{w_0}\right)^{|\ell|}L_\rho^{|\ell|}\!\left(\frac{2r'^2}{w_0^2}\right)e^{-\frac{r'^2}{w_0^2}}$. With $1/i^{\ell+n}=e^{-i\frac{\pi}{2}(\ell+n)}$, we obtain:
\begin{equation}
\begin{aligned}
E_{n,\ell,m}^{x,y}(r,\phi,z)&=2\pi\frac {e^{i\varphi_0(z)}}{i\lambda z}e^{\frac {ik}{2z}r^2}C_\rho^{|\ell|}\frac{1}{w_0}\left(\frac{\sqrt{2}}{w_0}\right)^{|\ell|}\cdot\alpha_{n,m}^{x,y}e^{i(\ell+n)(\phi-\frac{\pi}{2})}\cdot\int_0^{+\infty }dr'r'^{|\ell|+1}L_\rho^{|\ell|}\!\left(\frac{2r'^2}{w_0^2}\right)e^{-\frac{r'^2}{w_0^2}} J_{\ell+n}\left(\frac {krr'}{z}\right)=\\&=D_\rho^{|\ell|}(r,z)\alpha_{n,m}^{x,y}e^{i(\ell+n)(\phi-\pi/2)}H_{ n,\ell}(kr,z)
\label{eq:Fresnel4}
\end{aligned}
\end{equation}
\end{widetext}
where we introduced the function $D_\rho^{|\ell|}(r,z)=2\pi\frac {e^{i\varphi_0(z)}}{i\lambda z}e^{\frac {ik}{2z}r^2}C_\rho^{|\ell|}\frac{1}{w_0}\left(\frac{\sqrt{2}}{w_0}\right)^{|\ell|}$, and the function $H_{ n,\ell}(kr,z)$ as the result of the radial integral. This integral can be evaluated numerically, or for example it is found tabulated for the mode $\rho=0$ (meaning $L_\rho^{|\ell|}=1$) in Ref.~\cite{Rosenheinrich:2019}. 
In particular, we notice the fact that the $H$ functions are real, and that since $J_{-n}(x)=(-1)^n J_n(x)$ we have:
\begin{equation}
H_{-n,-\ell}(kr,z)=(-1)^{n+\ell}H_{n,\ell}(kr,z)
\label{eq:symh}
\end{equation}
From Eq.~\eqref{eq:Fresnel4} we confirm that the population of the $\ell+n$ modes is maintained after propagation, as expected.

\subsection{Expressions of the helicoidal dichroism}
The intensity detected on the screen placed in the far field, for each of the polarization components $x$ and $y$, will be the square modulus of the sum over $n$ of the field components $E_{n,\ell,m}^{x,y}$:
\begin{equation}
\begin{aligned}
	I_{\ell,m}^{x,y}&=\left|\sum_{n}E_{n,\ell,m}^{x,y}\right|^2=\\
	&=\left|D_\rho^{|\ell|}\right|^2
		\left|\sum_{n}\alpha_{n,m}^{x,y}e^{i(\ell+n)(\phi-\pi/2)} H_{n,\ell}\right|^2=\\
	&=\left|D_\rho^{|\ell|}\right|^2
		\sum_{n,n'} \alpha_{n,m}^{x,y} \overline{\alpha_{n',m}^{x,y}} e^{i(n-n')(\phi-\pi/2)}  H_{n,\ell}H_{n',\ell}=\\
	&=\left|D_\rho^{|\ell|}\right|^2
		\sum_{n} \left|\alpha_{n,m}^{x,y}\right|^2 H_{n,\ell}^2+\\
	&+
	\left|D_\rho^{|\ell|}\right|^2\sum_{n\neq n'}  \alpha_{n,m}^{x,y} \overline{\alpha_{n',m}^{x,y}} e^{i(n-n')(\phi-\pi/2)}  H_{n,\ell}H_{n',\ell}=\\
	&=\left(I_{\ell,m}^{x,y}\right)_1+\left(I_{\ell,m}^{x,y}\right)_2
\label{eq:Isimplified3}
\end{aligned}
\end{equation}
where we introduced the two terms of the sum $(I_{\ell,m}^{x,y})_{1,2}$ for convenience. Also, for the following it is useful to separate the product $\alpha_{n,m}^{x,y} \overline{\alpha_{n',m}^{x,y}}$ in its modulus and phase term:
\begin{equation}
    \alpha_{n,m}^{x,y} \overline{\alpha_{n',m}^{x,y}}=|\alpha_{n,m}^{x,y}||{\alpha_{n',m}^{x,y}}|e^{i \delta\varphi_{n,n'}^{x,y}}.
    \label{eq:defalphas2}
\end{equation}

At this point, we can explicitly write the expressions for MHD, where we need to calculate the difference in far-field intensities between two measurements. We will consider separately three possible dichroism experiments: incoming beams with opposite $\ell$ (MHD-$\ell$), magnetic targets with opposite magnetization direction $m$ (MHD-$m$), or both (MHD-$\ell m$).
We define the three MHD respectively as:
\begin{subequations}
\begin{align}
\Delta I^{x,y}_\ell(r,\phi,z)&=I_{\ell,m}^{x,y}(r,\phi,z)-I_{-\ell,m}^{x,y}(r,\phi,z)\\
\Delta I^{x,y}_m(r,\phi,z)&=I_{\ell,m}^{x,y}(r,\phi,z)-I_{\ell,-m}^{x,y}(r,\phi,z)\\
\Delta I^{x,y}_{\ell,m}(r,\phi,z)&=I_{\ell,m}^{x,y}(r,\phi,z)-I_{-\ell,-m}^{x,y}(r,\phi,z)
\end{align}
\label{eq:HDs}
\end{subequations}

Now we consider the effect of the two terms $(I_{\ell,m}^{x,y})_{1,2}$ on MHD separately.
We show in Appendix~\ref{app:AppendixTerm1} that the first term $(I_{\ell,m}^{x,y})_{1}$ does not contribute to MHD-$m$ and has negligible contribution to MHD-$\ell$ and MHD-$\ell m$ in most practical cases. Therefore, in the following we will disregard it. Instead, $(I_{\ell,m}^{x,y})_{2}$ leads to MHD.
As shown in Appendix~\ref{app:AppendixHD}, upon manipulation of the indices we can calculate the three MHD expressions:
\begin{widetext}
\begin{multline}
\Delta I_{\ell}^{x,y}(r,\phi,z)=2\left|D_\rho^{|\ell|}\right|^2\sum_{\substack{n\neq n'\\n-n'>0}} H_{n,\ell}(kr,z) H_{n',\ell}(kr,z)\cdot\\\cdot\left[|\alpha_{n,m}^{x,y}||\alpha_{n',m}^{x,y}|\cos\left((n-n')(\phi-\pi/2)+\delta\varphi^{x,y}_{n,n'}\right)\right.-
\left.(-1)^{n+n'}|\alpha_{-n,m}^{x,y}||\alpha_{-n',m}^{x,y}|\cos\left((n-n')(\phi-\pi/2)-\delta\varphi^{x,y}_{-n,-n'}\right)\right]
\label{eq:Isimplified4}
\end{multline}
\begin{equation}
\Delta I_{m}^{x,y}(r,\phi,z)=4\left|D_\rho^{|\ell|}\right|^2\sum_{\substack{ n\neq 0}} |\alpha^{x,y}_{0,m}||\alpha^{x,y}_{n,m}|\cos\left[n(\phi-\pi/2)+\delta\varphi^{x,y}_{n,0}\right]H_{0,\ell}(kr,z)H_{n,\ell}(kr,z)
\label{eq:Isimplified4b}
\end{equation}
\begin{multline}
\Delta I_{\ell,m}^{x,y}(r,\phi,z)=2\left|D_\rho^{|\ell|}\right|^2\sum_{\substack{n\neq n'\\n-n'>0}} H_{n,\ell}(kr,z) H_{n',\ell}(kr,z)\cdot\\\cdot\left[|\alpha_{n,m}^{x,y}||\alpha_{n',m}^{x,y}|\cos\left((n-n')(\phi-\pi/2)+\delta\varphi^{x,y}_{n,n'}\right)+\chi_{n,n'}
(-1)^{n+n'}|\alpha_{-n,m}^{x,y}||\alpha_{-n',m}^{x,y}|\cos\left((n-n')(\phi-\pi/2)-\delta\varphi^{x,y}_{-n,-n'}\right)\right]
\label{eq:Isimplified4c}
\end{multline}
\end{widetext}
For the third expression MHD-$\ell m$ of Eq.~\eqref{eq:Isimplified4c} we defined the function $\chi_{n,n'}=1$ if $n=0$ or $n'=0$ and $-1$ otherwise.

From Eq.~\eqref{eq:Isimplified4b} we can draw an important conclusion about the relationship between magnetic structure and observation of MHD. For this, we need to consider that having an OAM with fixed $\ell$ leads to MHD-$m$ as long as $\Delta I_{m}^{x,y}(r,\phi,z)\neq0$. Since $\alpha_{0,m}^{x,y}$, the $H$ function and the cosine term cannot be identically zero everywhere in $r$ and $\phi$, the only requirement is to have at least one $\alpha_{n,m}^{x,y}$ non zero. Therefore we find that \textbf{there will be MHD whenever the magnetization is non homogeneous}. Here it is important to stress that we considered the non homogeneity to be  an azimuthal dependence of the magnetization direction with constant magnitude. However, a decomposition of a radial dependent magnetization in a manner similar to Eq.~\eqref{eq:decomp} and a amplitude varying magnetization in Eq.~\eqref{alphas} would also eventually lead to MHD for similar reasons.

Furthermore, Eq.~\eqref{eq:Isimplified4b} leads us to another important conclusion. \textbf{In the particular case of an incoming beam without OAM ($\ell=0$), there still exists differential MHD when switching the magnetization (MHD-$m$)}. As soon as the magnetization has some structure (as in the two cases of Fig.\ref{fig:examples}, for example) a linearly polarized Gaussian beam will populate different OAM modes after reflection. As an example, the magnetic vortex [Fig.\ref{fig:examples}~(b)] will populate the modes $\ell=\pm1$. Measuring the reflected beam profile with spatial resolution will allow to obtain dichroic images when changing the sign of the magnetization. This is an extension to the case of MCD in reflection by a magnetic domain \cite{chauleau2018}, since we find that linearly polarized light even wihtout OAM can still lead to a helicoidal dichroic signal (MHD-$m$).

The three results of Eqs.~\eqref{eq:Isimplified4},\eqref{eq:Isimplified4b},\eqref{eq:Isimplified4c} are the general expressions of MHD for OAM light with generic $\ell$ and $\rho$ modes and any given symmetry of magnetization, but they are not trivial. For a certain structure with no specific symmetry, any coefficient may exist with any phase and we do not anticipate any further simplification. Instead, they can greatly simplify when considering specific symmetric structures with respect to the center of the beam, or specific incoming polarizations. An example is given in the following Section.

\subsection{Example: P-polarized beam on a symmetric structure}\label{sec:example}
In order to have a better insight and simplify the expressions of MHD, we consider the case of a \textit{P-polarized incoming beam reflected by a highly symmetric magnetic structure with decomposition coefficients $n$ of the same parity for $m_l$ and $m_t$}, as defined in Section~\ref{VortexModel}. In other words, if the component $m_l$ corresponds to a given case of Table~\ref{tablesym}, the component $m_t$ has to be in the same one or in the diagonal one. Therefore we set $\epsilon_s=0$, and since $m_{*,n}=\overline{m_{*,-n}}$ (Section~\ref{VortexModel}) we have $|\alpha_{n,m}^{x,y}|=|\alpha_{-n,m}^{x,y}|$. We also define the two following phases:
\begin{subequations}
\begin{align}
    \varphi_{*,n}&=\arg\left(m_{*,n}\right)+\arg\left(m\right)\\
    \varphi_0^t&=\arg\left(r_0^t\right)\,.
\end{align}
\end{subequations}
We have the following properties: $\varphi_{*,n}=-\varphi_{*,-n}$, and either $\varphi_{*,n}=0\;\text{or}\;\pi$ for real coefficients or $\varphi_{*,n}=\pm \pi/2$ for imaginary coefficients [see Table~\ref{tablesym}]. For the $\alpha^x$ values we calculate:
\begin{equation}
 \delta\varphi^{x}_{n,n'} = 
 \begin{cases}
   \varphi_{t,n}-\varphi_{t,n'}=0\;\text{or}\;\pi & \text{if $n, n'\neq 0$,}\\
    -\varphi_{t,n'}-\varphi_0^t & \text{if $n=0$ ,}\\
    \varphi_{t,n}+\varphi_0^t & \text{if $n'=0$.}
    \end{cases}\label{eq:deltaphi}
\end{equation}
Instead, for the $\alpha^y$ values we calculate:
\begin{subequations}
    \begin{align}
    \alpha_{0,m}^y&=0 \\
    \delta\varphi_{n,n'}^{y}&=\varphi_{l,n}-\varphi_{l,n'}=0\;or\;\pi\; \text{if $n,n'\neq 0$.}
    \end{align}
\end{subequations}

Now we can evaluate the three MHD expressions for this specific case. The full calculations are detailed in Appendix~\ref{app:AppendixHDex}. It is found that the $y$ component in all cases $\Delta I_{\ell}^{y}$, $\Delta I_{m}^{y}$ and $\Delta I_{\ell,m}^{y}$, is identically zero, therefore MHD reduces to the $x$ component. The results are reported in the following, and illustrated in Fig.~\ref{FigDichroisms} for the case of a magnetic dot with two antiparallel domains as in Fig.~\ref{fig:examples}(a).

\subsubsection*{Expression of MHD-$\ell$}
The expression of $\Delta I_{\ell}^{x}(r,\phi,z)$ takes a compact form when considering that only either odd or even $n$ terms are present because of the symmetries invoked in this example.
For even terms we obtain:
\begin{multline}
\Delta I_{\ell}^{x}(r,\phi,z)=-4\left|D_\rho^{|\ell|}\right|^2H_{0,\ell}|\alpha^{x}_{0,m}|\sin\varphi_0^t\cdot\\
\cdot\sum_{\substack{n\neq 0\\n\text{ even}}} (-1)^{\frac{n}{2}}H_{n,\ell}|\alpha^{x}_{n,m}|\sin\left(n\phi+\varphi_{t,n}\right)
\label{eq:IsimplifiedEven}
\end{multline}
\begin{figure}[!htp]
\centering
\includegraphics[width=0.5\textwidth]{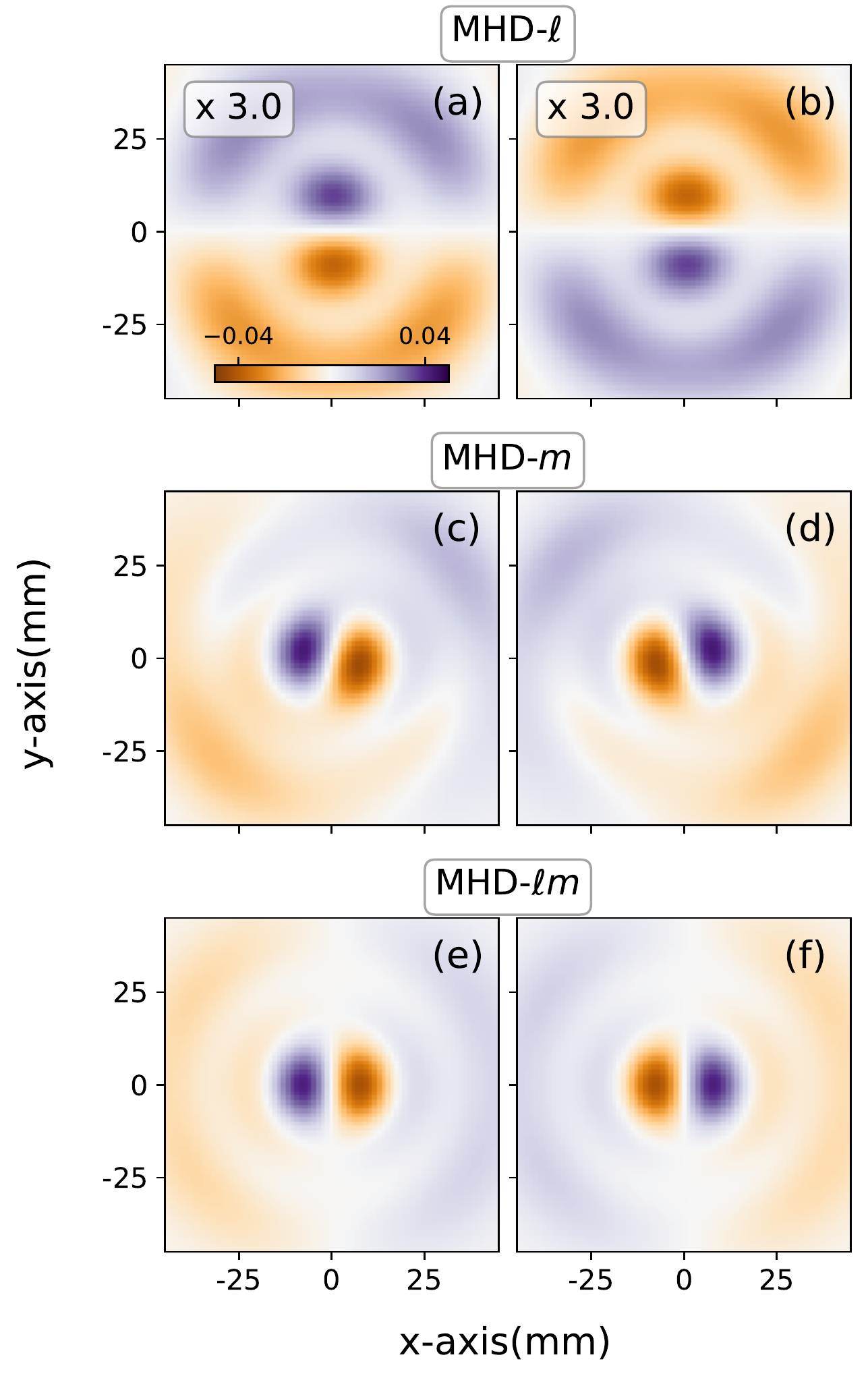}
\caption{
Computation of MHD for a magnetic dot of $500$~nm with two antiparallel domains as in Fig.~\ref{fig:examples}(a). (a) MHD-$\ell$ for $m=1$ and (b) for $m=-1$; (c) MHD-$m$ for $\ell=1$ and (d) for $\ell=-1$; (e) MHD-$\ell m$ as a difference of intensity maps corresponding to $\ell=1$ and $m=1$ and $\ell=-1$ and $m=-1$, and (f), MHD-$\ell m$, difference of intensity maps corresponding to $\ell=1$ and $m=-1$ and $\ell=-1$ and $m=1$. The MHD plots are normalized to the global maximum of their corresponding far-field intensity profiles. The incoming polarization is P and the angle of incidence is $\theta=5^\circ$. The computational details are given in Appendix~\ref{app:numerical}.
\label{FigDichroisms}}
 \end{figure}
while for odd terms we obtain:
\begin{multline}
\Delta I_{\ell}^{x}(r,\phi,z)=4\left|D_\rho^{|\ell|}\right|^2H_{0,\ell}|\alpha^{x}_{0,m}|\cos\varphi_0^t\cdot\\
\cdot\sum_{\substack{n\neq 0\\n\text{ odd}}}(-1)^{\frac{n+1}{2}} H_{n,\ell}|\alpha^{x}_{n,m}| \sin\left(n\phi+\varphi_{t,n}\right).
\label{eq:IsimplifiedOdd}
\end{multline}
In Fig.~\ref{FigDichroisms}(a),(b) we can observe this sine azimuthal dependence for the example of two antiparallel domains, calculated for $m=\pm1$.
\subsubsection*{Expression of MHD-$m$}
For the expression of $\Delta I_{m}^{x}(r,\phi,z)$ we obtain:
\begin{multline}
\Delta I_{m}^{x}(r,\phi,z)=4\left|D_\rho^{|\ell|}\right|^2H_{0,\ell}|\alpha^{x}_{0,m}|\cdot\\
\cdot\sum_{\substack{ n\neq 0}}H_{n,\ell}|\alpha^{x}_{n,m}|
\cos\left[n(\phi-\pi/2)+\varphi_{t,n}+\varphi_0^t\right]
\label{eq:HDMGeneral}
\end{multline}
We can observe this azimuthal dependence for the example of two antiparallel domains in Fig.~\ref{FigDichroisms}(c),(d), calculated for $\ell=\pm1$.
\subsubsection*{Expression of MHD-$\ell m$}
In a similar way to the case of MHD-$\ell$, the expression of $\Delta I_{\ell,m}^{x}(r,\phi,z)$ takes a compact form when considering separately the even and odd $n$ terms. 
For even terms we obtain:
\begin{multline}
\Delta I_{\ell, m}^{x}(r,\phi,z)=-4\left|D_\rho^{|\ell|}\right|^2H_{0,\ell}|\alpha^{x}_{0,m}|\cos\varphi_0^t\cdot\\
\cdot\sum_{\substack{n\neq 0\\n\text{ even}}} (-1)^{\frac{n}{2}}H_{n,\ell} |\alpha^{x}_{n,m}|\cos\left(n\phi+\varphi_{t,n}\right)
\label{eq:HD4Even}
\end{multline}
and for odd terms:
\begin{multline}
\Delta I_{\ell, m}^{x}(r,\phi,z)=4\left|D_\rho^{|\ell|}\right|^2H_{0,\ell}|\alpha^{x}_{0,m}|\sin\varphi_0^t\cdot\\
\cdot\sum_{\substack{n\neq 0\\n\text{ odd}}}(-1)^{\frac{n+1}{2}} H_{n,\ell} |\alpha^{x}_{n,m}|\cos\left(n\phi+\varphi_{t,n}\right)
\label{eq:HD4Odd}
\end{multline}
The cosine azimuthal dependence is shown in Fig.~\ref{FigDichroisms}(e),(f) for the example of two antiparallel domains, calculated for $m=\pm1$ or equivalently for $\ell=\pm1$.
\subsubsection*{Discussion of MHD expressions}
The five formulas Eqs.~\eqref{eq:IsimplifiedEven}-\eqref{eq:HD4Odd} summarize the main properties of MHD in structures with high symmetry as defined in Section~\ref{VortexModel}, probed with incoming linearly P-polarized light, and detected without any specific polarimetric device. We notice that all contributing terms come from the interference between regular reflectivity, indexed by the subscript 0, and a coefficient of the magnetization, with no cross terms between different magnetization components. We also note that inverting the magnetization corresponds to inverting all $\alpha^{x,y}_{n,m}$ coefficients ($n\neq0$), \textit{i.e.} adding a $\pi$-phase to all $\varphi_{t,n}$. All intensity differences change sign as expected for a dichroism. This is an important point if the process is to be used for determining the chirality of magnetic structures. Importantly, when the structure has only odd (resp. even) coefficients, $\Delta I_{\ell}^{x}(r,\phi,z)$ shows a sine pattern, the amplitude of which is proportional to $\cos\varphi_0^t$ (resp. $\sin\varphi_0^t$), while $\Delta I_{\ell,m}^{x}(r,\phi,z)$ shows a cosine pattern, the amplitude of which is proportional to $\sin\varphi_0^t$ (resp. $\cos\varphi_0^t$). The two patterns can thus be fitted to yield, up to some normalization the cosine and sine of the phase of one of the MOKE constants. In particular, if we consider the common case $|m_{t,n}||r_0^t|\ll1$, which is typical away from absorption resonance and Brewster's angle \cite{Valencia2006}, then $|\alpha_{n \neq 0,m}^x|\ll|\alpha_{0,m}^x|$ [Eq.~\eqref{alphas}], and therefore $I_\ell \propto |\alpha_{0,m}^x|^2 H_{0,\ell}^2 = |\epsilon_p r_{pp}|^2 H_{0,\ell}^2$, meaning that the reflected intensity is dominated by $|r_{pp}|^2$. In such a case one can estimate, e.g. for a sample with odd coefficients, the expression of the normalized dichroism $\text{MHD-}\ell=\Delta I_\ell/\left(I_\ell+I_{-\ell}\right)$:
\begin{multline}
\text{MHD-}\ell\approx2\sum_{\substack{n\neq 0\\n\; odd}}(-1)^{\frac{n+1}{2}} \frac{H_{n,\ell}(kr,z)}{H_{0,\ell}(kr,z)}\cdot\\
\cdot|r_0^t||m_{t,n}| \sin\left(n\phi+\varphi_{t,n}\right)\cos\varphi_0^t,
\label{eq:IsimplifiedOddHD}
\end{multline}
where we have reintroduced the explicit dependence of the $H_{n,\ell}$ functions on space and dropped the $x$ label, since there is strictly no contribution of the $y$ term to $\Delta I_{\ell}=(I_\ell-I_{-\ell})$, and only a weak one to $(I_\ell+I_{-\ell})$, of the order of the one discarded for the $x$-term. Similar expressions of the normalized MHD can be obtained in the other cases, replacing the sine and cosine according to Eqs. \eqref{eq:IsimplifiedEven},\eqref{eq:HD4Even} and \eqref{eq:HD4Odd}. 
Moreover, each component of the $\cos n\phi$ or $\sin n\phi$ images can be separated by a polar Fourier transform or an Abel inversion, and are all proportional to the amplitude of the MOKE constant times the $n$-th coefficient of the decomposition of the magnetization. We expect for this amplitude an order of magnitude corresponding to the ratio of the MOKE constants over the Fresnel constants, which is in the few percent to tens of percents.


\section{Conclusion}\label{sec:conclusion}
As allowed by Curie's principle, we identified a dichroism in the intensity pattern of beams carrying orbital angular momentum ($\ell$) reflected by a target with a magnetic structure, that we call Magnetic Helicoidal Dichroism. We have restricted our detailed analysis to the case of a magnetization distribution of constant amplitude, no radial dependence and with only in-plane components, but our model can be readily generalized beyond all these constraints. We find that as soon as the magnetization is not spatially homogeneous the reflected beam will populate different OAM modes depending on the decomposition of the magnetic structure on the polar basis set.
Consequently, the intensity pattern in the far field changes when changing $\ell$, the sign of the magnetization or both. The dichroism is differential in the sense that the effect is averaged out when the reflected intensity is integrated in space.
For magnetic structures of sufficiently high symmetry the MHD appears as a simple sinusoidal pattern at any given radius of the reflected image, the parameters of which depend on the particular shape of the magnetic structure and on the magneto-optical constants. In particular, we could directly link the differential image shapes to the coefficients of the polar decomposition of the magnetic structure, providing a new way to analyze the magnetization, or alternatively to determine the MOKE constants. For structures with many decomposition coefficients, the MOKE constants are even overdetermined and therefore reliably accessible in an experiment. This provides MHD with the potential of becoming an important new tool to access the properties of magnetic materials, in particular for dynamic studies measurements, which are known to be time-consuming, including time-resolved pump-probe studies of the magnetization dynamics that can be naturally implemented at laser based experimental facilities. 

The common situation of an incoming beam without OAM is a special case in our MHD model. We showed that reflection by a sample with inhomogeneous magnetization redistributes the mode $\ell=0$ into other modes with $\ell\neq0$, depending on the sample symmetry. The MHD is still present when switching the sign of the magnetization (MHD-$m$). This effect, up to now overlooked in magnetic reflection experiments, may provide a new way to either study magnetism or even to produce OAM light.

For practical reasons, an experiment exploiting the MHD requires reflection at an angle out from normal incidence. We found that the symmetry breaking of a tilted sample is an additional way to redistribute the OAM modes population upon reflection, which in general can be mixed to the redistribution due to the interaction with the structure. However, we cauld identify favorable conditions where the two effects do not mix. A specific case is an incoming P polarized beam, impinging on a structure that shows even or odd magnetization with respect to both the $x_\sigma$ and $y_\sigma$ axes. In this case, the magnetic effect only appears on the odd OAM modes, while the geometrical effect appears on the even ones. With this convenient separation in mind, we explore the particularly simple case of OAM light impinging on a magnetic vortex in the joint publication Ref.~\cite{MHDprl}. In fact, in such a case the required symmetry is respected and the decomposition has only two coefficients.

The theory of MHD that we have presented is very general and can be applied to many kinds of magnetic structures, to any polarization state and in different wavelength ranges. We developed it into details and applied it to a few specific cases, showing for instance how it makes it possible to retrieve the MOKE constants without implying any polarization analysis.  
The panel of possible applications, though, is much richer. To mention a few, one can envisage to: 
target resonance wavelengths to add element selectivity;
probe periodic nano- and micro-structures to address the collective response of their magnetization;
achieve a rapid readout of the magnetization state of structure, possibly with micrometer spatial resolution by integrating a beam and/or sample scanning system;
extend the MHD to the high photon energy and the time-resolved domains, intrinsically available with high harmonic generation sources and free electron lasers;
study the magnetization dynamics driven by the interaction with an OAM beam, because the redistribution of azimuthal modes may locally alter the magnetic structure of the sample, although the process occurs without any net transfer of angular momentum. 
From this list of outlooks, it appears that the analysis of MHD in the reflection of OAM beams from a magnetic structure can find original applications of both applied and fundamental interest in the field of magneto-optics.

\section*{Acknowledgements} We are very grateful to Giovanni De Ninno for fruitful discussions. 
This work was supported by the Agence Nationale pour la Recherche (under Contracts No. ANR11-EQPX0005-ATTOLAB and No. ANR14-CE320010-Xstase) and by the Swiss National Science Foundation project No. P2ELP2\_181877.

\clearpage
\appendix
\section{Numerical calculations}
\label{app:numerical}
\subsection{Field propagation}
The theory presented here, was tested with a specifically written python-based propagation code based on Fresnel propagation of light. The parameters of the code described below are shared with the joint publication Ref.~\cite{MHDprl} dedicated to the specific case of magnetic vortices. When propagating from the far-field to the near field or \emph{vice-versa}, a Fourier Transform method is used to compute the field in the new plane, while for far-field to far-field propagation we use the convolution method. 
We set Laguerre mask and lens at the same place (Fig.~\ref{fig:manip}), with focal length of the lens of $1$~m. The sample is placed at focus. The waist of the incoming collimated beam is set to $w_0=15$~mm. The intensity patterns are computed separately for the P and S linearly polarized components, before being combined when required.  The transverse direction is mapped on a spatial grid of 1024x1024 points. The wavelength of the beam is set to $\lambda=23.5$~nm when not specified otherwise. However, it is important to stress that the dependence of our results on the wavelength is only due to the variation of the reflectivity coefficients on $\lambda$, as illustrated in the last figure of Ref.~\cite{MHDprl}.

The two orthogonal components of the field calculated on the detector in the far field can be decomposed on a LG basis with a finite number of elements. In order to define the family of LG modes we need to choose the finite amount of $\rho$ and $\ell$ modes. We noted, on the examples that we treated, that using $\rho$ modes ranging from 0 to 18 and $\ell$ modes from -9 to 9 (that is 19 $\ell$ modes in total, and a grand total of 361 modes) was sufficient to accurately describe the intensity profiles. 
The decomposition on the LG basis is obtained by computing the integral of the product of any polarization component of the field ($S$ or $P$) with a given LG mode. If the result is below a given accuracy parameter (set here to $10^{-4}$)  compared to the dominant mode then it is forced to zero.
The trivial radius of curvature of the field wave front in the the far field, due to its divergence,  is removed before decomposition.

\subsection{Sample}
\begin{figure}[ht]
\centerline{\includegraphics[width=0.5\textwidth]{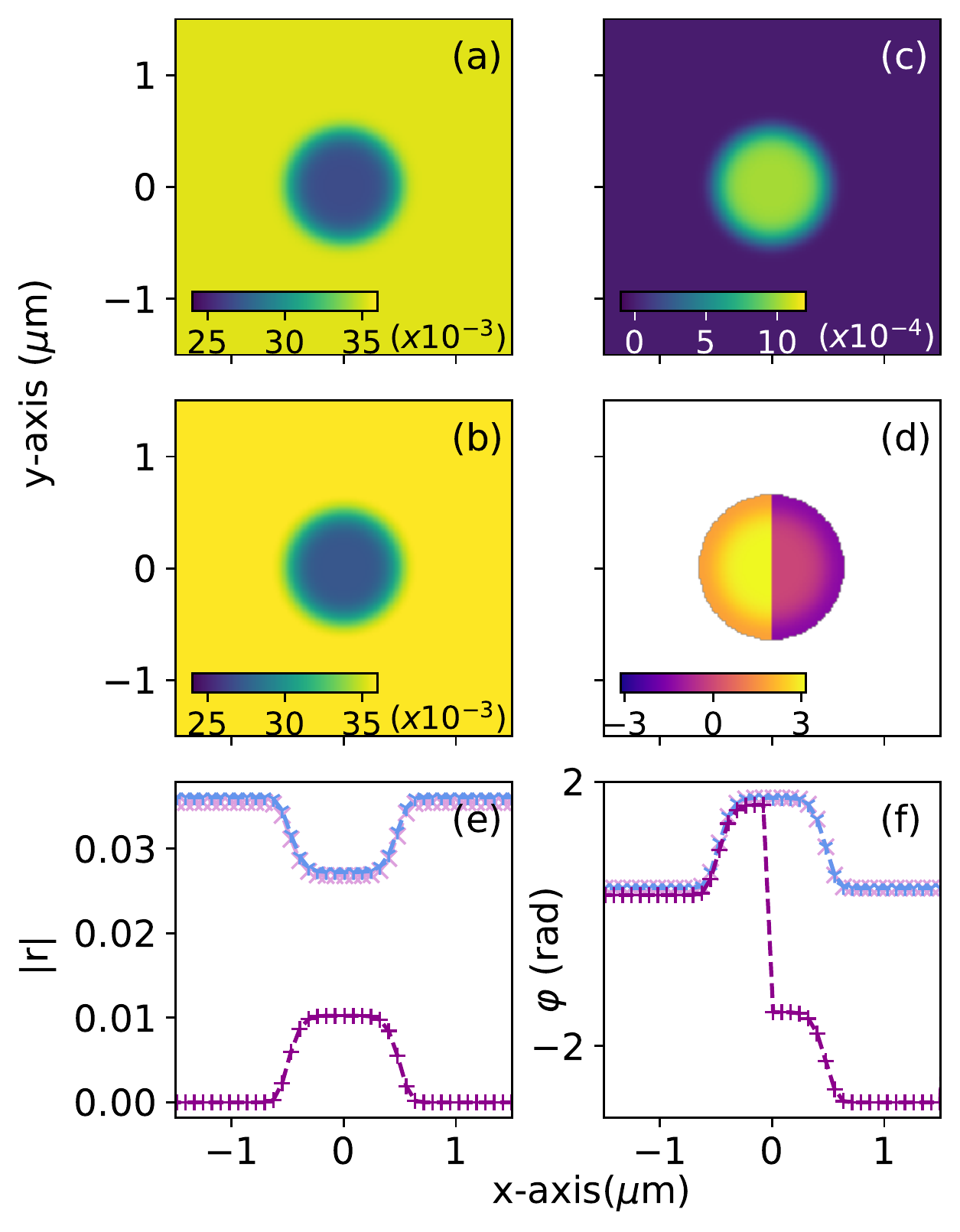}}
\caption{\label{Mapsreflectivity}(color online). Reflectivity coefficients maps used for the computation. $\lambda=23.5$\,nm, angle of incidence $\theta=5^\circ$.  From (a) to (d): $|r_{pp}|$, $|r_{ss}|$, $|r_{pp}\cdot r_0^t\cdot m_y|$ and $\arg(r_{pp}\cdot r_0^t\cdot m_y)$. For (d), all points with magnitudes of less than 1\%  of the  value at center (maximum), have been set to NaN. (e) Modulus and (f) phase of the following coefficients along the line at $y=0$. Plum cross: $r_{pp}$; blue dotted line: $r_{ss}$; purple plus: $r_0^t$.}
\end{figure}

In all numerical examples, we considered a SiN substrate (non magnetic material) on which the magnetic material, considered as Fe, was deposited. The optical reflectivity coefficients outside of the sample are thus those of SiN and not of a purely absorbing material, as used for sake of simplification in the analytical derivation in the main text. The corresponding $r_{pp}^\sigma$ and $r_{ss}^\sigma$ substrate values are displayed in Fig \ref{Mapsreflectivity}. This computational detail has of course no importance when the beam focus is smaller than the magnetic dot. 

In order to describe the magnetic dot used as a typical example here and in the joint publication Ref.~\cite{MHDprl}, we write its height $h$ as:
\begin{equation}
\label{eq:hypergauss}
h(r_\sigma)=h_0 e^{-\left(\frac{r_\sigma}{R_0}\right)^\beta}
\end{equation}
where $R_0$ is the radius of the dot, $h_0$ is the height of the dot in the center and $\beta$ is an integer. The hypergaussian function takes into account the sharpness of the dot edge.
In all numerical applications we set $\beta=6$.
The values of the reflectivity coefficients $r_{pp}$, $r_{ss}$, $r_0^t$ and $r_{ps}^l$ are imported from literature for Fe \cite{Valencia2006,ali1999,Zak1990,Piovera2013}, and are shown in Fig.~\ref{Mapsreflectivity} for the case of a magnetic dot of radius $R_0=500$~nm with two antiparallel magnetic domains as in the example of Fig.~\ref{fig:examples}(a). A smooth transition from the dot to the substrate is ensured by multiplying the reflectivity coefficients by $h(r_\sigma)/h_0$ [Eq. \eqref{eq:hypergauss}] for the magnetic dot and by $(1-h(r_\sigma))/h_0$ for the substrate. The reflectivity matrix thus reads
\begin{equation}
\begin{aligned}
\mathbf{R}(r_\sigma,\phi_\sigma)&=e^{\frac{i4\pi}{\lambda}h(r_\sigma)}\cdot \frac{h(r_\sigma)}{h_0}\cdot\\
&\begin{pmatrix}
r_{pp}\cdot\left[1+ r_0^t \cdot m_t(\phi_\sigma)\right]& r_{ps}^l \cdot m_l(\phi_\sigma) \\
- r_{ps}^l\cdot m_l(\phi_\sigma) & r_{ss}
\end{pmatrix}+\\
&\left(1-\frac{h(r_\sigma)}{h_0}\right)\cdot 
\begin{pmatrix}
r_{pp}^\sigma& 0 \\
0& r_{ss}^\sigma
\end{pmatrix}
\end{aligned}
\label{eq:reflectionMatrix2A}
\end{equation}
where we have explicitly indicated the dependence of the matrix on the azimuthal location on the sample. The leading phase term of the first line describes the dephasing of the light wave depending on whether it hits the dot or the substrate. For $\lambda\approx h_0$, this phase term is simply $\approx2\times2\pi$, where the factor $2$ describes back and forth travel. This dephasing is far from being negligible: the thickness of a permalloy dot is typically $h_0 \approx 20$~nm, comparable to the wavelength of the $3p$ resonance of Fe ($\lambda= 23.5$~nm).

\section{Parity of the angular decomposition}\label{AppendixSym}
We show here how to obtain the symmetry properties of Table~\ref{tablesym}. We express the coefficient $m_{*,n}$ by separating the azimuthal integral in Eq.~\ref{eq:decomp2} piecewise as:
\begin{equation}
\begin{aligned}
m_{*,n}&=\frac{1}{2\pi}\int_0^{\pi} m_*(\phi_\sigma)e^{in\phi_\sigma}d\phi_\sigma+\int_\pi^{2\pi} m_*(\phi_\sigma)e^{in\phi_\sigma}d\phi_\sigma\\
&=\frac{1}{2\pi}\int_0^{\pi}\left[ m_*(\phi_\sigma)e^{in\phi_\sigma}+m_*(2\pi-\phi_\sigma)e^{-in\phi_\sigma}\right]d\phi_\sigma\\
&=\frac{1}{2\pi}\int_0^{\frac{\pi}{2}}\left[ m_*(\phi_\sigma)e^{in\phi_\sigma}+m_*(2\pi-\phi_\sigma)e^{-in\phi_\sigma}\right]d\phi_\sigma+\\
& \qquad\int_{\frac{\pi}{2}}^{\pi}\left[ m_*(\phi_\sigma)e^{in\phi_\sigma}+m_*(2\pi-\phi_\sigma)e^{-in\phi_\sigma}\right]d\phi_\sigma\\
&=\frac{1}{2\pi}\int_0^{\frac{\pi}{2}}\left[ m_*(\phi_\sigma)e^{in\phi_\sigma}+m_*(2\pi-\phi_\sigma)e^{-in\phi_\sigma}\right]+\\
&(-1)^n \left[m_*(\pi-\phi_\sigma)e^{-in\phi_\sigma}+ m_*(\pi+\phi_\sigma)e^{in\phi_\sigma}\right]d\phi_\sigma
\label{eq:Generalcoeff}
\end{aligned}
\end{equation}
where we have used the changes of variables $\phi_\sigma\rightarrow2\pi-\phi_\sigma$ and $\phi_\sigma\rightarrow\pi-\phi_\sigma$ for the last integrals in lines 2 and 4 respectively.
We envision 4 general cases, depending on the symmetry of the azimuthal dependence with respect to $x_\sigma$ (i.e. $\phi_\sigma=\pi$) and $y_\sigma$ (i.e. $\phi_\sigma=\pi/2$) being even or odd, as listed in Table~\ref{tablesym}.

For instance, we consider $m_*$ being $(\text{even},\text{odd})$ w.r.t. $(x_\sigma,y_\sigma)$, i.e. $m_*(2\pi-\phi_\sigma)=m_*(\phi_\sigma)$ and $m_*(\pi-\phi_\sigma)=-m_*(\phi_\sigma)$. For example, this is the case of $m_t$ of both the antiparallel magnetic domains and the magnetic vortex of Fig.~\ref{fig:examples}. We have:
\begin{equation}
m_{*,n}= \frac{1}{2\pi}\int_0^{\frac{\pi}{2}} 2 m_*(\phi_\sigma)\cos n\phi_\sigma\left[1-(-1)^{n}\right]d\phi_\sigma\,.
\label{eq:evenodd}
\end{equation}
From this integral we see that all coefficients $m_{*,n}$ with even $n$ will be zero. Instead, they may be non zero for $n$ odd. In addition, all $m_{*,n}$ are real quantities, and $m_{*,-n}=m_{*,n}$.

An equivalent analysis for the other three symmetry cases leads to all the results presented in Table~\ref{tablesym}.

\section{Fourier decomposition of the magnetization of examples in Fig.~\ref{fig:examples}}
\label{app:magneticexamples}
We present the calculation of the Fourier decomposition coefficients for the magnetization in the case of antiparallel magnetic domains [Fig.~\ref{fig:examples}~(a)] and magnetic vortex [Fig.~\ref{fig:examples}~(b)].

\subsubsection*{Case of the antiparallel domains}
\begin{eqnarray*}
	m_{t,n} &=& \frac{m_0}{2\pi} \int_{-\pi}^{+\pi} d\phi_\sigma \text{sign}(\cos(\phi_\sigma)) e^{-in\phi_\sigma}\\
    &=& \frac{m_0}{2\pi} \left(
		\int_{-\pi/2}^{\pi/2} d\phi_\sigma e^{-in\phi_\sigma}
		- \int_{\pi/2}^{-\pi/2} d\phi_\sigma e^{-in\phi_\sigma}
	\right)\\
	&=& \frac{m_0}{2\pi} \frac{1}{-in} \Big[
		(-i)^n - i^n
		- \Big(i^n - (-i)^n \Big)
	\Big]\\
	&=& \frac{m_0 i^{n-1}}{n\pi}\Big[1-(-1)^n\Big]\\
m_{t,n} &=& \left\{\begin{array}{ll}
        0, & \text{for  n even}\\
       \frac{2 m_0 }{n \pi}i^{n-1}, & \text{for n odd}
        \end{array}\right.
\end{eqnarray*}

\subsubsection*{Case of the vortex}
For this case we use the general result:
\begin{equation}
\int_{-\pi}^{+\pi} d\phi_\sigma e^{i\kappa\phi_\sigma} = \left\{\begin{array}{ll}
        0, & \text{for }\kappa\neq0\\
       2\pi, & \text{for }\kappa=0
        \end{array}\right.
\end{equation}
We thus have for the longitudinal part:
\begin{equation}
m_{l,n} = \frac{m_0}{4i\pi}\int_{-\pi}^{+\pi} d\phi_\sigma (e^{-i\phi_\sigma} - e^{i\phi_\sigma}) e^{-in\phi_\sigma}
\end{equation}

\begin{equation}
m_{l,n} = \left\{\begin{array}{ll}
        0, & \text{for n $\neq \pm 1$}\\
       \mp \frac{m_0}{2i}, & \text{for $n = \pm 1$}
        \end{array}\right.
\end{equation}

and for the transverse part:
\begin{equation}
m_{t,n} = \frac{m_0}{4\pi}\int_{-\pi}^{+\pi} d\phi_\sigma (e^{i\phi_\sigma} + e^{-i\phi_\sigma}) e^{-in\phi_\sigma}
\end{equation}

\begin{equation}
m_{t,n} = \left\{\begin{array}{ll}
        0, & \text{for n $\neq \pm 1$}\\
       \frac{m_0}{2}, & \text{for $n = \pm 1$}
        \end{array}\right.
\end{equation}

\section{Trigonometry}
\label{sec:trigonometry}
In the following, we present the calculations that allow to obtain Eq.~\eqref{eq:trigonometry}. For the $\sin\phi_\sigma$ term:
\begin{subequations}
\begin{align}
\begin{split}
	\sin \phi_\sigma&=\frac{y_\sigma}{\sqrt{y_\sigma^2+x_\sigma^2}}\nonumber\\			
	&=\frac{y}{\sqrt{y^2+\frac{x^2}{\cos^2\theta}}}\nonumber\\
	&=\frac{y}{\sqrt{x^2+y^2}}\cdot\frac{1}{\sqrt{\frac{x^2/\cos^2\theta+y^2}{x^2+y^2}}}\nonumber\\
	&=\sin\phi\cdot\frac{1}{\sqrt{\frac{\cos^2\phi}{\cos^2\theta}+\sin^2\phi}}\nonumber\\
	&=\sin\phi\cdot\frac{\cos\theta}{\sqrt{\cos^2\phi+\sin^2\phi\cos^2\theta}}\nonumber\\
	&=\sin\phi\cdot\frac{\cos\theta}{\sqrt{1-\sin^2\phi\sin^2\theta}}\label{sineA}
\end{split}
\end{align}
\end{subequations}
The exact same computations can be carried out for the cosine, replacing the numerator of the first equation by $x_\sigma$. We thus get 
\begin{equation*}
	\cos \phi_\sigma=\cos\phi\cdot\frac{1}{\sqrt{1-\cos^2\phi\sin^2\theta}}
\end{equation*}
and combining them:
\begin{equation}
\begin{split}
	e^{i \phi_\sigma}&=\cos\phi_\sigma+i\sin\phi_\sigma\nonumber\\
	&=g(\phi,\theta)\cdot\left(\cos\phi +i\sin\phi \cos\theta\right)\nonumber\\
	&=g(\phi,\theta)\cdot\left(\frac{1+\cos\theta}{2}e^{i\phi} +\frac{1-\cos\theta}{2}e^{-i\phi}\right)\label{expA}
\end{split}
\end{equation}
For the $r_\sigma$ term we calculate:
\begin{equation}
\begin{split}
	r_\sigma&=\sqrt{y_\sigma^2+x_\sigma^2}\nonumber\\			
	&=\sqrt{y^2+\frac{x^2}{\cos^2\theta}}\nonumber\\
	&=\sqrt{x^2+y^2}\cdot\sqrt{\frac{x^2/\cos^2\theta+y^2}{x^2+y^2}}\nonumber\\
	&=r\cdot\sqrt{\frac{\cos^2\phi}{\cos^2\theta}+\sin^2\phi}\nonumber\\
	&=\frac{r}{\cos\theta}\sqrt{1-\sin^2\phi\sin^2\theta}\label{rA}
\end{split}
\end{equation}

\section{Contribution of $(I_{\ell,m}^{x,y})_{1}$ to MHD}\label{app:AppendixTerm1}
In this Appendix we show that the contribution of the term $(I_{\ell,m}^{x,y})_{1}$ to MHD is negligible.

\subsubsection*{Case of MHD-$\ell$}
We calculate $(\Delta I_{\ell})_{1}=(I_{\ell,m}^{x,y})_{1}-(I_{-\ell,m}^{x,y})_{1}$:

\begin{eqnarray*}
	(\Delta I_{l})_{1}
		&\propto& 
			\left|\alpha_{0,m}^{x,y}\right|^2
			(H_{0,\ell}^2 - H_{0,-\ell}^2)+\\
		&+& \sum_{n>0}
			\left|\alpha_{n,m}^{x,y}\right|^2
			(H_{n,\ell}^2 - H_{n,-\ell}^2)+\\
		&+& \sum_{n<0}
			\left|\alpha_{n,m}^{x,y}\right|^2
			(H_{n,\ell}^2 - H_{n,-\ell}^2)\\
		&\propto&
			\sum_{n>0}
			(\left|\alpha_{n,m}^{x,y}\right|^2 - \left|\alpha_{-n,m}^{x,y}\right|^2)
			(H_{n,\ell}^2 - H_{n,-\ell}^2)
\end{eqnarray*}
Now, considering Eq.~\eqref{alphas}, for the $y$ component we see that since $\left|\alpha_{n,m}^{y}\right|^2 = \left|\alpha_{-n,m}^{y}\right|^2$, we have $(\Delta I_{\ell})_{1}=0$. 

Instead, for the $x$ component we have $\alpha_{n,m}^{x}$ defined as the sum of two terms. If one of them is zero, for instance the incoming light is P- or S-polarized ($\epsilon_s=0$ or $\epsilon_p=0$, respectively) or if the magnetic structure is such that $m_{t,n}$ or $m_{l,n}$ are zero, then also in this case we have simply $\left|\alpha_{n,m}^{x}\right|^2 = \left|\alpha_{-n,m}^{x}\right|^2$, and thus $(\Delta I_{\ell})_{1}=0$.

If the experimental conditions are such that both components of $\alpha_{n,m}^{x}$ are not zero, then $(\Delta I_{\ell})_{1}$ is not necessarily zero. A somewhat lengthy calculation leads to:
\begin{eqnarray*}
(\left|\alpha_{n,m}^{x}\right|^2 - \left|\alpha_{-n,m}^{x}\right|^2)=-4|r_0^t|^2|\epsilon_p\epsilon_sr_{pp}r_{ps}^l|\cdot\\ \cdot\text{Im}\{m_{t,n}\overline{m_{l,n}}\}\sin(\varphi_p-\varphi_s+\varphi_{pp}-\varphi_{ps}^l)\,,
\end{eqnarray*}
namely a term proportional to the square of the magnetization and of the MOKE constant $r_0^t$. This can be an important contribution compared to $(\Delta I_{\ell})_{2}$ in the most general case, and we note that it could be a way to experimentally access the quantity $\varphi_{pp}-\varphi_{ps}^l$ under properly choosen conditions. However, as we will see later (Section~\ref{sec:example}), for highly symmetric structures and under the reasonable assumption $|m_{*,n}||r_0^t|\ll1$, $(\Delta I_{\ell})_{2}$ is found to be linear with magnetization and $r_0^t$, thus $(\Delta I_{\ell})_{1}$ can still be considered negligible. In any case, as pointed out above this issue can be simply avoided by choosing a proper experimental setup.

\subsubsection*{Caseof MHD-$m$}
We calculate $(\Delta I_{m})_{1}=(I_{\ell,m}^{x,y})_{1}-(I_{\ell,-m}^{x,y})_{1}$:
\begin{equation}
(\Delta I_{m})_{1}\propto\sum_n(\left|\alpha_{n,m}^{x,y}\right|^2-\left|\alpha_{n,-m}^{x,y}\right|^2)H_{n,\ell}^2\,.
\end{equation}
Since $\left|\alpha_{n,m}^{x,y}\right|^2 = \left|\alpha_{n,-m}^{x,y}\right|^2$, in this case simply $(\Delta I_{m})_{1}$ is always zero.

\subsubsection*{Case of MHD-$\ell m$}
A similar analysis to that for MHD-$\ell$ leads to the same conclusion of $(\Delta I_{\ell,m})_{1}$ giving a negligible contribution to MHD-$\ell m$. 

\clearpage
\begin{widetext}
\section{Derivation of general MHD formulas}\label{app:AppendixHD}
In the following we present the detailed derivation of the general MHD expressions, where we evaluate the three differences of Eq.~\eqref{eq:HDs} using the expression of $(I_{\ell,m}^{x,y})_{2}$ from Eq.~\eqref{eq:Isimplified3}. In order to simplify the notation we drop the $(kr,z)$ explicit dependence of the $H_{n,l}$ and the $m$ subscript for all $\alpha_{n}^{x,y}$ terms. We will use the property $H_{-n,-\ell}=(-1)^{n+\ell}H_{n,\ell}$ [Eq.~\eqref{eq:symh}], and the general property of series $\sum_{n-n'>0}s_{n,n'}=\sum_{n'-n>0}s_{n',n}=\sum_{n-n'>0}s_{-n',-n}$.

\subsection*{Expression of MHD-$\ell$ [Eq.~\eqref{eq:Isimplified4}]}
\begin{equation}
\begin{aligned}
\Delta I_{\ell}^{x,y}(r,\phi,z)&=\left|D_\rho^{|\ell|}\right|^2\sum_{n\neq n'} \left(\alpha^{x,y}_{n}\overline{\alpha^{x,y}_{n'}}e^{i(n-n')(\phi-\pi/2)} \right)   \left[H_{n,\ell} H_{n',\ell}-H_{n,-\ell} H_{n',-\ell}\right]=\\
&=2\left|D_\rho^{|\ell|}\right|^2\sum_{\substack{n\neq n'\\n-n'>0}} |\alpha^{x,y}_{n}||\alpha^{x,y}_{n'}|\cos\left[(n-n')(\phi-\pi/2)+\delta\varphi_{n,n'}^{x,y}\right]    \left[H_{n,\ell} H_{n',\ell}-H_{n,-\ell} H_{n',-\ell}\right]=\\
&=2\left|D_\rho^{|\ell|}\right|^2\sum_{\substack{n\neq n'\\n-n'>0}} |\alpha^{x,y}_{n}||\alpha^{x,y}_{n'}|\cos\left[(n-n')(\phi-\pi/2)+\delta\varphi_{n,n'}^{x,y}\right]    \left[H_{n,\ell} H_{n',\ell}-(-1)^{n+n'}H_{-n,\ell} H_{-n',\ell}\right]=\\
&=2\left|D_\rho^{|\ell|}\right|^2\sum_{\substack{n\neq n'\\n-n'>0}} |\alpha^{x,y}_{n}||\alpha^{x,y}_{n'}|\cos\left[(n-n')(\phi-\pi/2)+\delta\varphi_{n,n'}^{x,y}\right]    H_{n,\ell} H_{n',\ell}+\\
&\qquad-2\left|D_\rho^{|\ell|}\right|^2\sum_{\substack{n\neq n'\\n-n'>0}}(-1)^{n+n'}|\alpha^{x,y}_{n}||\alpha^{x,y}_{n'}|\cos\left[(n-n')(\phi-\pi/2)+\delta\varphi_{n,n'}^{x,y}\right]
H_{-n,\ell} H_{-n',\ell}=\\
&=2\left|D_\rho^{|\ell|}\right|^2\sum_{\substack{n\neq n'\\n-n'>0}} |\alpha^{x,y}_{n}||\alpha^{x,y}_{n'}|\cos\left[(n-n')(\phi-\pi/2)+\delta\varphi_{n,n'}^{x,y}\right]    H_{n,\ell} H_{n',\ell}+\\
&\qquad-2\left|D_\rho^{|\ell|}\right|^2\sum_{\substack{n\neq n'\\-n+n'>0}}(-1)^{n+n'}|\alpha^{x,y}_{-n}||\alpha^{x,y}_{-n'}|\cos\left[-(n-n')(\phi-\pi/2)+\delta\varphi_{-n,-n'}^{x,y}\right]
H_{n,\ell} H_{n',\ell}=\\
&=2\left|D_\rho^{|\ell|}\right|^2\sum_{\substack{n\neq n'\\n-n'>0}} |\alpha^{x,y}_{n}||\alpha^{x,y}_{n'}|\cos\left[(n-n')(\phi-\pi/2)+\delta\varphi_{n,n'}^{x,y}\right]    H_{n,\ell} H_{n',\ell}+\\
&\qquad-2\left|D_\rho^{|\ell|}\right|^2\sum_{\substack{n\neq n'\\n-n'>0}}(-1)^{n+n'}|\alpha^{x,y}_{-n'}||\alpha^{x,y}_{-n}|\cos\left[-(n'-n)(\phi-\pi/2)+\delta\varphi_{-n',-n}^{x,y}\right]
H_{n',\ell} H_{n,\ell}=\\
\Delta I_{\ell}^{x,y}(r,\phi,z)&=2\left|D_\rho^{|\ell|}\right|^2\sum_{\substack{n\neq n'\\n-n'>0}} H_{n,\ell} H_{n',\ell}\left[|\alpha^{x,y}_{n}||\alpha^{x,y}_{n'}|\cos\left((n-n')(\phi-\pi/2)+\delta\varphi_{n,n'}^{x,y}\right)\right.-\\
&\qquad\qquad\qquad\left.(-1)^{n+n'}|\alpha^{x,y}_{-n}||\alpha^{x,y}_{-n'}|\cos\left((n-n')(\phi-\pi/2)-\delta\varphi_{-n,-n'}^{x,y}\right)\right]
\label{eq:IsimplifiedAppendix}
\end{aligned}
\end{equation}

\subsection*{Expression of MHD-$m$ [Eq.~\eqref{eq:Isimplified4b}]}
\begin{equation}
\begin{aligned}
\Delta I_{m}^{x,y}(r,\phi,z)&=\left|D_\rho^{|\ell|}\right|^2\sum_{n\neq n'} \left(\alpha^{x,y}_{n}\overline{\alpha^{x,y}_{n'}}e^{i(n-n')(\phi-\pi/2)} \right)   \left[H_{n,\ell} H_{n',\ell}+\chi_{n,n'} H_{n,\ell} H_{n',\ell}\right]=\\
&=2\left|D_\rho^{|\ell|}\right|^2\sum_{\substack{n\neq n'\\n-n'>0}} |\alpha^{x,y}_{n}||\alpha^{x,y}_{n'}|\cos\left[(n-n')(\phi-\pi/2)+\delta\varphi_{n,n'}^{x,y}\right]    \left[H_{n,\ell} H_{n',\ell}+\chi_{n,n'}H_{n,\ell} H_{n',\ell}\right]=\\
&=4\left|D_\rho^{|\ell|}\right|^2\sum_{\substack{n= 0\; or \; n'= 0\\n-n'>0}} |\alpha^{x,y}_{n}||\alpha^{x,y}_{n'}|\cos\left[(n-n')(\phi-\pi/2)+\delta\varphi_{n,n'}^{x,y}\right]    H_{n,\ell} H_{n',\ell}=\\
&=4\left|D_\rho^{|\ell|}\right|^2\sum_{\substack{-n'>0}} |\alpha^{x,y}_{0}||\alpha^{x,y}_{n'}|\cos\left[-n'(\phi-\pi/2)+\delta\varphi_{0,n'}^{x,y}\right]    H_{0,\ell} H_{n',\ell}+\\
&\qquad\qquad +4\left|D_\rho^{|\ell|}\right|^2\sum_{\substack{n>0}} |\alpha^{x,y}_{n}||\alpha^{x,y}_{0}|\cos\left[n(\phi-\pi/2)+\delta\varphi_{n,0}^{x,y}\right]    H_{n,\ell} H_{0,\ell}=\\
&=4\left|D_\rho^{|\ell|}\right|^2\sum_{\substack{n<0}} |\alpha^{x,y}_{0}||\alpha^{x,y}_{n}|\cos\left[-n(\phi-\pi/2)-\delta\varphi_{n,0}^{x,y}\right]    H_{0,\ell} H_{n,\ell}+\\
&\qquad\qquad +4\left|D_\rho^{|\ell|}\right|^2\sum_{\substack{n>0}} |\alpha^{x,y}_{n}||\alpha^{x,y}_{0}|\cos\left[n(\phi-\pi/2)+\delta\varphi_{n,0}^{x,y}\right]    H_{n,\ell} H_{0,\ell}=\\
&=4\left|D_\rho^{|\ell|}\right|^2\sum_{\substack{n\neq 0}} |\alpha^{x,y}_{0}||\alpha^{x,y}_{n}|\cos\left[n(\phi-\pi/2)+\delta\varphi_{n,0}^{x,y}\right]    H_{0,\ell} H_{n,\ell}=\\
\Delta I_{m}^{x,y}(r,\phi,z)&=4\left|D_\rho^{|\ell|}\right|^2|\alpha^{x,y}_{0}|H_{0,\ell}\sum_{\substack{n\neq 0}} |\alpha^{x,y}_{n}|\cos\left[n(\phi-\pi/2)+\delta\varphi_{n,0}^{x,y}\right] H_{n,\ell}
\label{eq:IsimplifiedAppendix2}
\end{aligned}
\end{equation}

\subsection*{Expression of MHD-$\ell m$ [Eq.~\eqref{eq:Isimplified4c}]}
The derivation of the expression of $\Delta I_{\ell,m}^{x,y}(r,\phi,z)$ follows the same steps as that of $\Delta I_{\ell}^{x,y}(r,\phi,z)$ [Eq.~\eqref{eq:Isimplified4}].

\section{Derivation of MHD formulas for the example of Section~\ref{sec:example}}\label{app:AppendixHDex}
In the following we drop the explicit dependence of the $H$ functions on the spatial coordinates and the $m$ subscript for all $\alpha^{x,y}_n$ terms. Also, it is useful to express differences and sums of the $\delta\varphi^x$ phase terms as:
\begin{subequations}
\begin{align}
    &\frac{\delta\varphi^{x}_{n,n'}-\delta\varphi^{x}_{-n,-n'}}{2}=\begin{cases}
    \varphi_{t,n}-\varphi_{t,n'}=0\;\text{or}\;\pi& \text{if $n,n'\neq 0$,}\\
    -\varphi_{t,n'} & \text{if $n=0$ ,}\\
    \varphi_{t,n}  & \text{if $n'=0$.}
    \end{cases}\\
    &\frac{\delta\varphi^{x}_{n,n'}+\delta\varphi^{x}_{-n,-n'}}{2}= \begin{cases}
    0\;
    &\text{if $n,n'\neq 0$,}\\
    -\varphi_0^t& \text{if $n=0$ ,}\\
    \varphi_0^t  & \text{if $n'=0$.}
    \end{cases}
    \end{align}
\end{subequations}

\subsection*{Expression of MHD-$\ell$ [Eqs.~\eqref{eq:IsimplifiedEven},\eqref{eq:IsimplifiedOdd}]}
Taking into account that $|\alpha_{n,m}^{x,y}|=|\alpha_{-n,m}^{x,y}|$ and using prosthaphaeresis identity, the MHD-$\ell$ from Eq.~\eqref{eq:Isimplified4} reads:
\begin{multline}
\begin{aligned}
\Delta I_{\ell}^{x,y}(r,\phi,z)&=\\
-4\left|D_\rho^{|\ell|}\right|^2&\sum_{\substack{n\neq n'\\n-n'>0\\n+n'\text{ even}}} (-1)^{\frac{n-n'}{2}}H_{n,\ell} H_{n',\ell}|\alpha_{n}^{x,y}||\alpha_{n'}^{x,y}| \sin\left((n-n')\phi+\frac{\delta\varphi_{n,n'}^{x,y}-\delta\varphi_{-n,-n'}^{x,y}}{2}\right)\sin\left(\frac{\delta\varphi_{n,n'}^{x,y}+\delta\varphi_{-n,-n'}^{x,y}}{2}\right)+\\
+4\left|D_\rho^{|\ell|}\right|^2&\sum_{\substack{n\neq n'\\n-n'>0\\n+n'\text{ odd}}}(-1)^{\frac{n-n'+1}{2}} H_{n,\ell} H_{n',\ell}|\alpha^{x,y}_{n}||\alpha^{x,y}_{n'}| \sin\left((n-n')\phi+\frac{\delta\varphi_{n,n'}^{x,y}-\delta\varphi_{-n,-n'}^{x,y}}{2}\right)\cos\left(\frac{\delta\varphi_{n,n'}^{x,y}+\delta\varphi_{-n,-n'}^{x,y}}{2}\right)
\label{eq:Isimplified5}
\end{aligned}
\end{multline}
However, for the $y$ term, we notice that the first line in the expression of $\Delta I_{\ell}^{y}(r,\phi,z)$ is zero due to the final sine when $n,n'\neq 0$ and to $\alpha_{0,m}^y=0$ when $n=0$ or $n'=0$. For the second line, if both $n\neq 0$ and $n' \neq 0$, in both cases of symmetries yielding either only odd or only even terms, the sum $n+n'$ is even and is not in the sum. If $n= 0$ or $n'= 0$, $\alpha_{0,m}^y=0$ and the contribution is null. Therefore, in the framework of our approximations we identically have $\Delta I_{\ell}^{y}(r,\phi,z)=0$. 

Instead, the $x$ term $\Delta I_{\ell}^{x}(r,\phi,z)$ can be further simplified considering that only either odd or even terms are present because of the symmetries invoked in this example.

\subsubsection*{Even terms}
For even terms, only the first line of  Eq.~\eqref{eq:Isimplified5} is present since $n+n'$ is necessarily even. If both $n\neq 0$ and $n' \neq 0$, however, the sum is zero because of the last sine term. We are thus left with the sum over either $n$ or $n'=0$. Since the sum runs over $n-n'>0$, it amounts to summing, for all even integers, the same expression. We thus split the sum in two sub-sums: 
\begin{equation}
\begin{aligned}
\Delta I_{\ell}^{x}(r,\phi,z)&=
-4\left|D_\rho^{|\ell|}\right|^2\sum_{\substack{n\neq 0\\n>0\\n\text{ even}}} (-1)^{\frac{n}{2}}H_{n,\ell} H_{0,\ell}|\alpha^{x}_{n}||\alpha^{x}_{0}| \sin\left(n\phi+\frac{\delta\varphi_{n,0}^{x}-\delta\varphi_{-n,0}^{x}}{2}\right)\sin\left(\frac{\delta\varphi_{n,0}^{x}+\delta\varphi_{-n,0}^{x}}{2}\right)+\\
&\qquad\qquad-4\left|D_\rho^{|\ell|}\right|^2\sum_{\substack{0\neq n'\\-n'>0\\n'\text{ even}}} (-1)^{\frac{-n'}{2}}H_{0,\ell} H_{n',\ell}|\alpha^{x}_{0}||\alpha^{x}_{n'}| \sin\left(-n'\phi+\frac{\delta\varphi_{0,n'}^{x}-\delta\varphi_{0,-n'}^{x}}{2}\right)\sin\left(\frac{\delta\varphi_{0,n'}^{x}+\delta\varphi_{0,-n'}^{x}}{2}\right)=\\
&=-4\left|D_\rho^{|\ell|}\right|^2\sum_{\substack{n\neq 0\\n>0\\n\text{ even}}} (-1)^{\frac{n}{2}}H_{n,\ell} H_{0,\ell}|\alpha^{x}_{n}||\alpha^{x}_{0}| \sin\left(n\phi+\varphi_{t,n}\right)\sin\left(\varphi_0^t\right)+\\
&\qquad\qquad-4\left|D_\rho^{|\ell|}\right|^2\sum_{\substack{n\neq 0\\n<0\\n\text{ even}}} (-1)^{\frac{n}{2}}H_{n,\ell} H_{0,\ell}|\alpha^{x}_{n}||\alpha^{x}_{0}| \sin\left(-n\phi-\varphi_{t,n}\right)\sin\left(-\varphi_0^t\right)=\\
&=-4\left|D_\rho^{|\ell|}\right|^2\sum_{\substack{n\neq 0\\n\text{ even}}} (-1)^{\frac{n}{2}}H_{n,\ell} H_{0,\ell}|\alpha^{x}_{n}||\alpha^{x}_{0}| \sin\left(n\phi+\varphi_{t,n}\right)\sin\left(\varphi_0^t\right).
\label{eq:IsimplifiedAppEven}
\end{aligned}
\end{equation}
finding the result of Eq.~\eqref{eq:IsimplifiedEven}.

\subsubsection*{Odd terms}
For odd terms, the first line of  Eq.~\eqref{eq:Isimplified5} is present only if both $n\neq 0$ and $n' \neq 0$, however the sum is zero because of the last sine term, and thus only the second line needs to be considered. This also reduces to a sum over either $n=0$ or $n'=0$, since it is the only way to have $n+n'$ odd:
\begin{equation}
\begin{aligned}
\Delta I_{\ell}^{x}(r,\phi,z)&=4\left|D_\rho^{|\ell|}\right|^2\sum_{\substack{n\neq 0\\n>0\\n\text{ odd}}}(-1)^{\frac{n+1}{2}} H_{n,\ell} H_{0,\ell}|\alpha^{x}_{n}||\alpha^{x}_{0}| \sin\left(n\phi+\frac{\delta\varphi_{n,0}^{x}-\delta\varphi_{-n,0}^{x}}{2}\right)\cos\left(\frac{\delta\varphi_{n,0}^{x}+\delta\varphi_{-n,0}^{x}}{2}\right)+\\
&+4\left|D_\rho^{|\ell|}\right|^2\sum_{\substack{0\neq n'\\-n'>0\\n'\text{ odd}}}(-1)^{\frac{-n'+1}{2}} H_{0,\ell} H_{n',\ell}|\alpha^{x}_{0}||\alpha^{x}_{n'}| \sin\left(-n'\phi+\frac{\delta\varphi_{0,n'}^{x}-\delta\varphi_{0,-n'}^{x}}{2}\right)\cos\left(\frac{\delta\varphi_{0,n'}^{x}+\delta\varphi_{0,-n'}^{x}}{2}\right)=\\
&=4\left|D_\rho^{|\ell|}\right|^2\sum_{\substack{n\neq 0\\n>0\\n\text{ odd}}}(-1)^{\frac{n+1}{2}} H_{n,\ell} H_{0,\ell}|\alpha^{x}_{n}||\alpha^{x}_{0}| \sin\left(n\phi+\varphi_{t,n}\right)\cos\left(\varphi_0^t\right)+\\
&\qquad\qquad+4\left|D_\rho^{|\ell|}\right|^2\sum_{\substack{0\neq n'\\n<0\\n\text{ odd}}}(-1)^{\frac{-n+1}{2}} H_{0,\ell} H_{n,\ell}|\alpha^{x}_{0}||\alpha^{x}_{n}| \sin\left(-n\phi-\varphi_{t,n}\right)\cos\left(-\varphi_0^t\right)=\\
&=4\left|D_\rho^{|\ell|}\right|^2\sum_{\substack{n\neq 0\\n\text{ odd}}}(-1)^{\frac{n+1}{2}} H_{n,\ell} H_{0,\ell}|\alpha^{x}_{n}||\alpha^{x}_{0}| \sin\left(n\phi+\varphi_{t,n}\right)\cos\left(\varphi_0^t\right)
\label{eq:IsimplifiedAppOdd}
\end{aligned}
\end{equation}
where we have taken into account for the last step that $(-1)^{\frac{-n+1}{2}}=(-1)^{\frac{n-1}{2}}=-(-1)^{\frac{n+1}{2}}$, and we find the result of Eq.~\eqref{eq:IsimplifiedOdd}.

\subsection*{Expression of MHD-$m$ [Eq.~\eqref{eq:HDMGeneral}]}
The expression of Eq.~\eqref{eq:HDMGeneral} is simply obtained by using the definition of $\delta\varphi_{n,0}^x$ from Eq.~\eqref{eq:deltaphi} into Eq.~\eqref{eq:Isimplified4b}.

\subsection*{Expression of MHD-$\ell m$ [Eqs.~\eqref{eq:HD4Even},\eqref{eq:HD4Odd}]}
Similarly to the case of MHD-$\ell$, we can calculate the full expression of MHD-$\ell m$. Following a similar derivation as for Eq.~\eqref{eq:Isimplified5}, the expression of $\Delta I_{\ell,m}^{x,y}(r,\phi,z)$ is:
\begin{multline}
\begin{aligned}
\Delta I_{\ell,m}^{x,y}(r,\phi,z)&=\\
4\left|D_\rho^{|\ell|}\right|^2&\sum_{\substack{n\neq n'\\n-n'>0\\n+n'\text{ even}\\n=0\, or \, n'=0}} (-1)^{\frac{n-n'}{2}}H_{n,\ell} H_{n',\ell}|\alpha^{x,y}_{n}||\alpha^{x,y}_{n'}| \cos\left((n-n')\phi+\frac{\delta\varphi_{n,n'}^{x,y}-\delta\varphi_{-n,-n'}^{x,y}}{2}\right)\cos\left(\frac{\delta\varphi_{n,n'}^{x,y}+\delta\varphi_{-n,-n'}^{x,y}}{2}\right)+\\
-4\left|D_\rho^{|\ell|}\right|^2&\sum_{\substack{n\neq n'\\n-n'>0\\n+n'\text{ odd}\\n=0\, or \, n'=0}}(-1)^{\frac{n-n'+1}{2}} H_{n,\ell} H_{n',\ell}|\alpha^{x,y}_{n}||\alpha^{x,y}_{n'}| \cos\left((n-n')\phi+\frac{\delta\varphi_{n,n'}^{x,y}-\delta\varphi_{-n,-n'}^{x,y}}{2}\right)\sin\left(\frac{\delta\varphi_{n,n'}^{x,y}+\delta\varphi_{-n,-n'}^{x,y}}{2}\right)
\label{eq:Isimplified5c}
\end{aligned}
\end{multline}
At this point, similarly to the case of MHD-$\ell$, we can separate the calculations into the case of even or odd $n$ terms.

\subsubsection*{Even terms}
\begin{equation}
\begin{aligned}
\Delta I_{\ell,m}^{x}(r,\phi,z)&=
-4\left|D_\rho^{|\ell|}\right|^2\sum_{\substack{n\neq 0\\n>0\\n\text{ even}}} (-1)^{\frac{n}{2}}H^{(n,\ell)} H_{0,\ell}|\alpha^{x}_{n}||\alpha^{x}_{0}| \cos\left(n\phi+\frac{\delta\varphi_{n,0}^{x}-\delta\varphi_{-n,0}^{x}}{2}\right)\cos\left(\frac{\delta\varphi_{n,0}^{x}+\delta\varphi_{-n,0}^{x}}{2}\right)+\\
&\qquad\qquad-4\left|D_\rho^{|\ell|}\right|^2\sum_{\substack{0\neq n'\\-n'>0\\n'\text{ even}}} (-1)^{\frac{-n'}{2}}H_{0,\ell} H_{n',\ell}|\alpha^{x}_{0}||\alpha^{x}_{n'}| \cos\left(-n'\phi+\frac{\delta\varphi_{0,n'}^{x}-\delta\varphi_{0,-n'}^{x}}{2}\right)\cos\left(\frac{\delta\varphi_{0,n'}^{x}+\delta\varphi_{0,-n'}^{x}}{2}\right)=\\
&=-4\left|D_\rho^{|\ell|}\right|^2\sum_{\substack{n\neq 0\\n>0\\n\text{ even}}} (-1)^{\frac{n}{2}}H_{n,\ell} H_{0,\ell}|\alpha^{x}_{n}||\alpha^{x}_{0}| \cos\left(n\phi+\varphi_{t,n}\right)\cos\left(\varphi_0^t\right)+\\
&\qquad\qquad-4\left|D_\rho^{|\ell|}\right|^2\sum_{\substack{n\neq 0\\n<0\\n\text{ even}}} (-1)^{\frac{n}{2}}H_{n,\ell} H_{0,\ell}|\alpha^{x}_{n}||\alpha^{x}_{0}| \cos\left(-n\phi-\varphi_{t,n}\right)\cos\left(-\varphi_0^t\right)=\\
&=-4\left|D_\rho^{|\ell|}\right|^2\sum_{\substack{n\neq 0\\n\text{ even}}} (-1)^{\frac{n}{2}}H_{n,\ell} H_{0,\ell}|\alpha^{x}_{n}||\alpha^{x}_{0}| \cos\left(n\phi+\varphi_{t,n}\right)\cos\left(\varphi_0^t\right)
\label{eq:IsimplifiedAppEvenC}
\end{aligned}
\end{equation}
finding the result of Eq.~\eqref{eq:HD4Even}.

\subsubsection*{Odd terms}
\begin{equation}
\begin{aligned}
\Delta I_{\ell,m}^{x}(r,\phi,z)&=4\left|D_\rho^{|\ell|}\right|^2\sum_{\substack{n\neq 0\\n>0\\n\text{ odd}}}(-1)^{\frac{n+1}{2}} H_{n,\ell} H_{0,\ell}|\alpha^{x}_{n}||\alpha^{x}_{0}| \cos\left(n\phi+\frac{\delta\varphi_{n,0}^{x}-\delta\varphi_{-n,0}^{x}}{2}\right)\sin\left(\frac{\delta\varphi_{n,0}^{x}+\delta\varphi_{-n,0}^{x}}{2}\right)+\\
&+4\left|D_\rho^{|\ell|}\right|^2\sum_{\substack{0\neq n'\\-n'>0\\n'\text{ odd}}}(-1)^{\frac{-n'+1}{2}} H_{0,\ell} H_{n',\ell}|\alpha^{x}_{0}||\alpha^{x}_{n'}| \cos\left(-n'\phi+\frac{\delta\varphi_{0,n'}^{x}-\delta\varphi_{0,-n'}^{x}}{2}\right)\sin\left(\frac{\delta\varphi_{0,n'}^{x}+\delta\varphi_{0,-n'}^{x}}{2}\right)=\\
&=4\left|D_\rho^{|\ell|}\right|^2\sum_{\substack{n\neq 0\\n>0\\n\text{ odd}}}(-1)^{\frac{n+1}{2}} H_{n,\ell} H_{0,\ell}|\alpha^{x}_{n}||\alpha^{x}_{0}| \cos\left(n\phi+\varphi_{t,n}\right)\sin\left(\varphi_0^t\right)+\\
&\qquad\qquad+4\left|D_\rho^{|\ell|}\right|^2\sum_{\substack{0\neq n\\n<0\\n\text{ odd}}}(-1)^{\frac{-n+1}{2}} H_{0,\ell} H_{n,\ell}|\alpha^{x}_{0}||\alpha^{x}_{n}| \cos\left(-n\phi-\varphi_{t,n}\right)\sin\left(-\varphi_0^t\right)=\\
&=4\left|D_\rho^{|\ell|}\right|^2\sum_{\substack{n\neq 0\\n\text{ odd}}}(-1)^{\frac{n+1}{2}} H_{n,\ell} H_{0,\ell}|\alpha^{x}_{n}||\alpha^{x}_{0}| \cos\left(n\phi+\varphi_{t,n}\right)\sin\left(\varphi_0^t\right)
\label{eq:IsimplifiedAppOddC}
\end{aligned}
\end{equation}
finding the result of Eq.~\eqref{eq:HD4Odd}.
\end{widetext}

\bibliographystyle{unsrt}
\end{document}